\documentclass[10pt,prl,twocolumn,preprint,nofootinbib]{revtex4}

\usepackage{epsfig}
\usepackage{amsmath,amsfonts,amssymb}
\usepackage{t1enc}
\usepackage{verbatim}
\usepackage{float}
\usepackage{morefloats}
\usepackage{booktabs}
\usepackage{slashed}
\usepackage{xcolor}
\begin{document}
\title{CP tests of Higgs couplings in \(t\bar{t}h\) semileptonic events at the LHC}

\author{
	D. Azevedo\(^{1,4}\), A. Onofre\(^2\), F. Filthaut\(^{3}\), R. Gon\c calo\(^4\)
	\\[3mm]
	{\footnotesize {\it
			$^1$ Department of Physics and Astronomy, Universidade do Porto,\\ Rua do Campo Alegre 1021/1055, 4169-007 Porto, Portugal\\
			$^2$ Departamento de F\'{\i}sica, Universidade do Minho, 4710-057 Braga, Portugal\\
			$^3$ Radboud University and Nikhef, Heyendaalseweg 135, 6525 AJ Nijmegen, The Netherlands\\
			$^4$ LIP, Av. Prof. Gama Pinto, n\({}^\circ\)2, 1649-003 Lisboa, Portugal\\
	}}
}

%%%%  Abstract  %%%%%%
\begin{abstract}
The CP nature of the Higgs coupling to top quarks is addressed in this paper, in single charged lepton final states of \(t\bar{t}h\) events produced in proton-proton collisions at the LHC. Pure scalar (\(h=H\)) and pseudo-scalar (\(h=A\)) Higgs boson signal events, generated with MadGraph5\_aMC@NLO, are fully reconstructed using a kinematic fit. Angular distributions of the decay products, as well as CP-sensitive asymmetries, are exploited to separate and gain sensitivity to possible pseudo-scalar components of the Higgs boson and reduce the contribution from the dominant irreducible background \(t\bar{t}b\bar{b}\). 
Significant differences are found between the pure CP-even and -odd signal hypotheses as well as with respect to the Standard Model background, in particular the  $t\bar{t} b\bar{b}$ contribution.
Such differences survive the event reconstruction, allowing to define optimal observables to extract the Higgs couplings parameters from a global fit. A dedicated analysis is applied to efficiently identify signal events and reject as much as possible the expected Standard Model background. The results obtained are compared with a similar analysis in the dilepton channel. We show that the single lepton channel is more promising overall and can be used in combination to study the CP nature of the Higgs coupling to top quarks.

\end{abstract}

\maketitle

%%%% Main Body %%%%%%
\section{I. Introduction}
A new particle with mass around 125 GeV, consistent with the Higgs boson predicted by the Standard Model (SM), was discovered by ATLAS~\cite{:2012gk} and CMS~\cite{:2012gu} at the Large Hadron Collider (LHC). This discovery is of fundamental importance for the electroweak symmetry breaking mechanism~\cite{higgsmech} which allows elementary particles to acquire mass. The properties of the Higgs boson have been extensively studied ever since and, in particular, its couplings. Even though the SM predictions for the Higgs boson are in remarkable agreement with experimental results~\cite{Khachatryan:2014kca,Khachatryan:2016tnr,Aad:2015mxa}, the SM cannot be the ultimate theory. It does not explain the baryon asymmetry in the Universe, which may require additional sources of CP violation. The SM also fails to provide a viable dark matter candidate. Extensions with multiple Higgs doublets~\cite{MultiHiggs} can provide new sources of CP violation in the Higgs sector. These may have an impact on the Higgs Yukawa couplings to fermions, by adding a new CP-odd component to the SM coupling.

Although a pure CP-odd case was already excluded at 99.98\% confidence level (CL) in $\gamma$, $Z$ and $W$ interactions with Higgs bosons~\cite{Khachatryan:2014kca,Khachatryan:2016tnr}, mixing between a CP-even and CP-odd component is allowed by experimental data. Moreover, the couplings of the Higgs boson to fermions might show a different nature where large CP-odd couplings are still allowed and may be different among the different flavors of fermions~\cite{Fontes:2017zfn}.
Of particular relevance, is the measurement of the Higgs couplings in associated production with top quarks. Since the Higgs Yukawa coupling to the top quark is expected to be close to unity~\cite{ttHtheory}, much larger than the other Yukawa couplings, its impact on the vacuum stability is expected to be more important. The study of this process allows a direct measurement of the vertex and, in particular, provides sensitivity to the CP nature of the Higgs couplings to top quarks~\cite{BhupalDev:2007ftb}. 
The precise evaluation of the total production cross section to the highest order in QCD is of crucial importance to study this rare process at the LHC. The impact of soft gluon resummation to next-to-next-to-leading logarithmic (NNLL) accuracy is already available~\cite{Broggio:2015lya, Broggio:2016lfj}.

In this paper the associated production of Higgs boson with a pair of top quarks ($t\bar t h$) is studied at the LHC for a centre-of-mass energy of 13 TeV. While semileptonic decays of the \(t\bar{t}\) system are searched for 
($t\bar t\rightarrow bW^+\bar{b}W^-\rightarrow b\bar{b}q\bar q'\ell^{\pm}\nu_{\ell}$), the Higgs is expected to decay through the SM dominant decay mode (\(H\to b\bar{b}\)). The single lepton final state topology is characterised by the presence of an isolated charged lepton and missing transverse energy (\( \slashed{E}\)) from the undetected neutrino. 

The observation of the associated production of top quarks with a Higgs boson was recently announced independently by the CMS and the ATLAS collaborations, using a combination of analyses, resulting in significances in excess of 5 standard deviations~\cite{Sirunyan:2018hoz, Aaboud:2018urx}. These analyses employed LHC data collected at 7, 8 and 13~TeV and are sensitive to $t\bar{t}h$ final states, with the Higgs boson decaying to $b\bar{b}$, $WW^*$, $\tau^+\tau^-$, $\gamma\gamma$ and $ZZ^*$. These observations open the field to the type of measurements proposed in the current paper.

The analysis of $t\bar{t}h$ ($h\rightarrow b\bar b$) final states at the LHC is particularly challenging due to the low expected cross section 
and large \(t\bar{t} + \text{jets}\) background. For the Higgs decay channel considered in this paper, $h\rightarrow b\bar b$, the \(t\bar{t}b\bar{b}\) background is of particular importance. 
We search for deviations from the SM by comparing the kinematics of \(t\bar{t}h\) signals with SM-like couplings (\(h=H\) and \(J^\text{CP}=0^+\)) with \(t\bar{t}h\) signals with pure pseudo-scalar Higgs bosons (\(h=A\) and \(J^\text{CP}=0^-\)). The most general Lagrangian that accounts for contributions from CP-even and CP-odd components of the couplings is defined as,
\begin{equation}\label{eq1}
\mathcal{L}=\kappa y_t \bar{t} (\cos\alpha+i\gamma_5 \sin\alpha)th
\end{equation}
where \(y_t\) is the SM Higgs boson Yukawa coupling and \(\alpha\) represents a CP phase. While the SM interaction is recovered by fixing \(|\cos\alpha|=1\), the pure pseudo-scalar is obtained by setting \(\cos\alpha=0\).

We consider several CP-sensitive angular distributions introduced in the literature~\cite{Gunion:1996xu, Boudjema:2015nda, Santos:2015dja}. These are evaluated for the first time in semileptonic final states of \(t\bar{t}h\) decays, in this paper. 
Full event kinematic reconstruction is applied to reconstruct the four-momenta of all massive particles ($t, \bar t, h, W^+$ and $W^-$) as well as the undetected neutrino. We show that, even after showering, detector simulation, event selection, and full kinematic reconstruction, the distributions of several angular variables are largely preserved. Moreover, background discrimination in this channel can be enhanced using these angular variables.
Although results are consistent with what was observed in an analysis of the dileptonic $t\bar{t}h$ channel~\cite{Santos:2015dja, AmorDosSantos:2017ayi}, the signal sensitivity of the semileptonic analysis reported here is larger than the dileptonic one.
As in the dileptonic analysis ~\cite{AmorDosSantos:2017ayi} all the mixed states of CP-even and CP-odd couplings gave results between the ones obtained with $\cos{\alpha}=$0 and $\cos{\alpha}=$1, only these two extreme cases are considered in the present analysis.

This paper is organised as follows. We begin with a brief introduction in Section I and a description of the event generation, simulation and event selection in Section II. In Section III we analise the angular variables, as reconstructed in the $t\bar{t}h$ semileptonic channel and, in Section IV, the results are discussed. The conclusions are summarised in Section V.

\section{II. Simulation and selection}

%%%%% Events Selection %%%%%%%
\vspace*{2mm}
{\it \noindent Monte Carlo Generation}
\vspace*{2mm}

The generation of the $t\bar{t}h$ signals (both scalar and pseudo-scalar) and the $t\bar t b \bar b$ dominant background were performed, at next-to-leading-order (NLO) in QCD, with Madgraph5\_aMC@NLO~\cite{Alwall:2014hca}. The NNPDF2.3 PDF sets~\cite{Ball:2012cx} were used. While the default model ({\scshape sm}) was used for the CP-even SM Higgs boson signal ($h=H$), the generation of the pure CP-odd pseudo-scalar signal ($h=A$), used the {\scshape hc\_nlo\_x0} model~\cite{Artoisenet:2013puc}. 
The Madgraph5\_aMC@NLO event-generator NLO cross sections were assumed for all three of these samples. For the case of the $t\bar{t}H$ signal this is in agreement with a recent calculation at approximate next-to-next-to-leading order (NNLO) accuracy including threshold resummation of soft gluon emission in the soft-collinear effective theory framework~\cite{Broggio:2015lya}. The best current knowledge of the $t\bar{t}H$ cross section comes from next-to-next-to-logarithmic (NNLL) calculations~\cite{Broggio:2016lfj}, which show only small diferences with respect to the NLO cross section but roughly factor two improved precision.
In addition to the $t\bar t b \bar b$ dominant background, other sources of SM backgrounds were also considered. These included \(t\bar{t}+\text{ jets}\) (where \textit{jets} stands for up to 3 additional jets from the hadronization of c- or light-flavoured quarks), \(t\bar{t}V+\text{ jets}\) (where \(V=W^\pm, Z\) and \textit{jets} can go up to one additional jet), single top quark production ($s$-channel, $t$-channel and $Wt$ associated production), diboson (\(W^+W^-, ZZ, W^\pm Z + \text{jets}\) with up to 3 additional jets), \(W^\pm+ \text{ jets}\)  (with up to 4 additional jets) and \(W b\bar{b}+\text{ jets}\) (with up to 2 additional jets). These backgrounds were also generated with MadGraph5\_aMC@NLO but at leading-order (LO) in QCD. For the generation at LO we used the {\scshape MLM}~\cite{Alwall:2007fs} matching scheme. 
The cross-section of the \(t\bar{t}+\text{jets}\) background was normalized to the next-to-next-to-leading order (NNLO) in QCD with next-to-next-to-leading-logarithm (NNLL) resummation of soft gluon terms~\cite{Czakon:2011xx,Botje:2011sn,Martin:2009bu,Gao:2013xoa,Ball:2012cx}. 
The electroweak single top quark production cross section was scaled to the approximate NNLO theoretical calculation~\cite{Kidonakis:2010tc,Kidonakis:2011wy}.  Although full NNLO calculations exist \cite{Brucherseifer:2014ama} for single top quark production, the approximate cross section was used instead and rescaled to the exact top quark mass used in the generation according to the prescription given in~\cite{Czakon:2013goa}. The same prescription was applied to the \(t\bar{t}+\text{jets}\) background. For all the other SM background processes, the MadGraph5\_aMC@NLO event-generator cross sections were used. The generation was performed at a centre-of-mass energy of 13~TeV, at the LHC, with dynamic renormalization and factorization scales set to the sum of the transverse masses of all final state partons. The masses of the top quark ($m_t$), the $W$ boson ($m_W$) and Higgs bosons (for both scalar, $m_H$, and pseudo-scalar, $m_A$) were set to $173$~GeV, $80.4$~GeV and $125$~GeV, respectively. In order to preserve full spin correlations, {\scshape MadSpin}~\cite{Artoisenet:2012st} was used to decay heavy particles. Parton showering and hadronization was perfomed by {\scshape Pythia6}~\cite{Sjostrand:2006za}. 

%%%%% Events Simulation %%%%%%%
\vspace*{2mm}
{\it \noindent Event Simulation and Reconstruction} 
\vspace*{2mm}

Following generation and parton showering, events were passed through a fast simulation of a typical LHC detector, using {\scshape Delphes}~\cite{deFavereau:2013fsa}. This allows to have more realistic experimental conditions in what concerns the reconstruction of charged leptons, jets and missing transverse energy (and missing momentum). The efficiencies and resolutions of the default detector subsystems are parametrized as a function of transverse momentum, $p_T$,
and pseudo-rapidity, $\eta$, for the different types of particles (details may be found in~\cite{deFavereau:2013fsa}). {\scshape{FastJet}}~\cite{Cacciari:2011ma} was used for jet reconstruction using the anti-$k_t$ algorithm~\cite{Cacciari:2008gp} with radial parameter R set to 0.4. The efficiency ($\epsilon_b$) for tagging jets originating from the hadronization of $b$ quarks, i.e. $b$-tagging, is dependent on their transverse momentum (\(p_T\)) according to,
\begin{equation}
	\epsilon_b(p_T)=0.8\tanh(0.003 p_T)\frac{30}{1+0.086p_T},
\end{equation}
with $p_{T}$ given in GeV, in the region where $p_T\geq10$~GeV and $\vert\eta\vert\leq2.5$. 
The efficiency is set to zero outside this region. 
The mis-tag probability of identifying light and $c$-jets as fake $b$-jets, is given by:
\begin{equation}
	\epsilon_m(p_T)=0.002+7.3\times10^{-6}p_T.
\end{equation}

The analysis of simulated events was performed with {\scshape MadAnalysis 5}~\cite{Conte:2012fm} in the expert mode~\cite{Conte:2014zja}. In order to build the angular distributions, kinematic properties of signal events need to be fully reconstructed.  This was accomplished by {\scshape KLFitter}~\cite{Erdmann:2013rxa} with the likelihood-based reconstruction method. {\scshape KLFitter} uses {\it transfer functions}, $W_k(E_k^{\text{meas}}|E_k^{\text{parton}})$, to reconstruct particle energies ($E_k^{\text{parton}}$) using their measured values ($E_k^{\text{meas}}$) after detector simulation, together with the knowledge of experimental resolutions. These are considered for jets ($k=j$) and charged leptons ($k=\ell$). We implemented a dedicated parametrization of the jet transfer functions, which depend on their energy and pseudo-rapidity.
We also applied the transfer function $W_{\text{miss}}(E_{\text{miss},x(y)}^{\text{meas}}|E_{\nu,x(y)}^{\text{parton}})$, to reconstruct the $x$($y$) component of the neutrino transverse energy $E_{\nu,x(y)}^{\text{parton}}$, from the measured $x$($y$) missing transverse energy component, $E_{\text{miss},x(y)}^{\text{meas}}$.  
To make sure the transfer functions were appropriate for semileptonic $t\bar th$ final states, we applied a {\it pre-selection} by requiring events to have, at least, 6 jets and one charged lepton. These cuts were applied in the definition of the transfer functions themselves. Once all were defined, the likelihood was built according to, 
\begin{equation}
\begin{aligned}
\hspace*{-15mm}
L & = B(m_{{b_{had}},q_1,q_2} | m_t, \Gamma_t) \times B(m_{q_1,q_2}| m_W,\Gamma_W) \\
               &\times B(m_{{b_{lep}},   \ell,  \nu} | m_t,\Gamma_t)  \times B(m_{ \ell, \nu} | m_W,\Gamma_W) \\ 
               &\times B(m_{{b_{h},\bar{b}_{h}}} | m_h, \Gamma_h)
\end{aligned}
\end{equation}
\vspace*{-8mm}
\begin{equation}
\begin{aligned}               
             ~~~~~ & \times  \prod_{i=1}^6 W_i^{\text{jet}}(E_i^{\text{meas}}| E^{\text{parton}}_i) \times W_\ell(E_\ell^{\text{meas}} | E_\ell^{\text{parton}}) \\
              & \times W_{\text{miss}}(E_{\text{miss},x}^{\text{meas}} | E_{\nu,x}^{\text{parton}}) \times W_{\text{miss}}(E_{\text{miss},y}^{\text{meas}} | E_{\nu,y}^{\text{parton}}) \nonumber\\
\end{aligned}
\label{eq:MLElikelihood}
\end{equation}
which used several Breit-Wigner probability density functions, \(B(m_{x_1,x_2,\dots}|m_{X},\Gamma_X )\), to evaluate the probability of reconstructing the invariant mass ($m_{x_1,x_2,\dots}$) of system ${x_1,x_2,\dots}$, consistent with a particle of mass $m_X$ and width 
$\Gamma_X$ ($X=t, W$ and $h$). We tried all possible permutations of all measured jets, in order to find the best matching between the jets and
\begin{itemize}
\item[i)]  the two $c$- or light-flavoured quarks ($q_1$,$q_2$), from the hadronically decaying \(W\) boson, as well as,
\item[ii)]  the two $b$-quarks from the decay of the Higgs boson ($b_h$, $\bar b_{h}$) and,
\item[iii)]  the two $b$-quarks, one from the fully hadronic ($b_{\text{had}}$) and the other from the semileptonic ($b_{\text{lep}}$) decays of the top quarks present in the events.
\end{itemize}

The lepton ($\ell$) and undetected neutrino ($\nu$) from the leptonically decaying \(W\) boson, together with the $b_{\text{lep}}$ candidate in each permutation, were used to reconstruct the top quark that decayed through the semileptonic decay.
The neutrino $p_z$ reconstruction was accomplished by considering the $x$ ($y$) components of the missing transverse energy to be the $x$ ($y$) components of the neutrino's momentum, constrained by
\begin{equation}
m_W^2=(p_\nu + p_\ell)^2.
\end{equation}
If, for any permutation, two solutions were found, the one that maximized the likelihood function was selected. When no solution was found for a particular permutation, the neutrino $p_z$ was fixed to zero.
From the long list of all possible permutations and solutions, the one chosen as the best candidate for the full kinematic reconstruction of the event was the one with the largest value of the likelihood function.
The partons' four-momenta were reconstructed from the objects of that particular permutation and, in order to accommodate the corrections from the transfer functions, their energy was changed to that obtained in the kinematic fit. The momentum components were also rescaled according to
\begin{equation}
\vec{p}_i^\text{ parton}=\xi_i \vec{p}_i^\text{ meas},
\end{equation}
with
\begin{equation}
\xi_i = \sqrt{\frac{(E^\text{\text{parton}}_i)^2 - m_i^2}{(E^\text{\text{meas}}_i)^2 - m_i^2}} . 
\end{equation}

%%%%% Events Selection %%%%%%%
\vspace*{2mm}
{\it \noindent Event Selection}
\vspace*{2mm}

Following the pre-selection and kinematic reconstruction, additional selection criteria were applied to events, defining what we call the {\it final selection}. Charged leptons and $b$-tagged jets were required to have \(p_T\geq20\)~GeV and \(|\eta|\leq2.5\). For non \(b\)-tagged jets the $\eta$ selection was increased to \(|\eta|\leq4.5\). Only events with  \( \slashed{E}\)>20~GeV were used, since the final state topology involves one undetected neutrino. Furthermore, only topologies with 6 to 8 jets, 3 or 4 of which were \(b\)-tagged, were considered in the event analysis. We checked these topologies have the largest matching efficiency ($\epsilon_{\text{match}}$=30.1\%), defined as the fraction of events for which all objects from the chosen permutation were within $\Delta R = \sqrt{(\Delta\eta)^2+(\Delta\phi)^2}$<0.4 of the corresponding partons at generator level. At this stage, all selected events were reconstructed by {\scshape KLFitter}. 

\begin{figure*}
	\vspace*{5cm}
\begin{center}
\begin{tabular}{ccc}
\epsfig{file=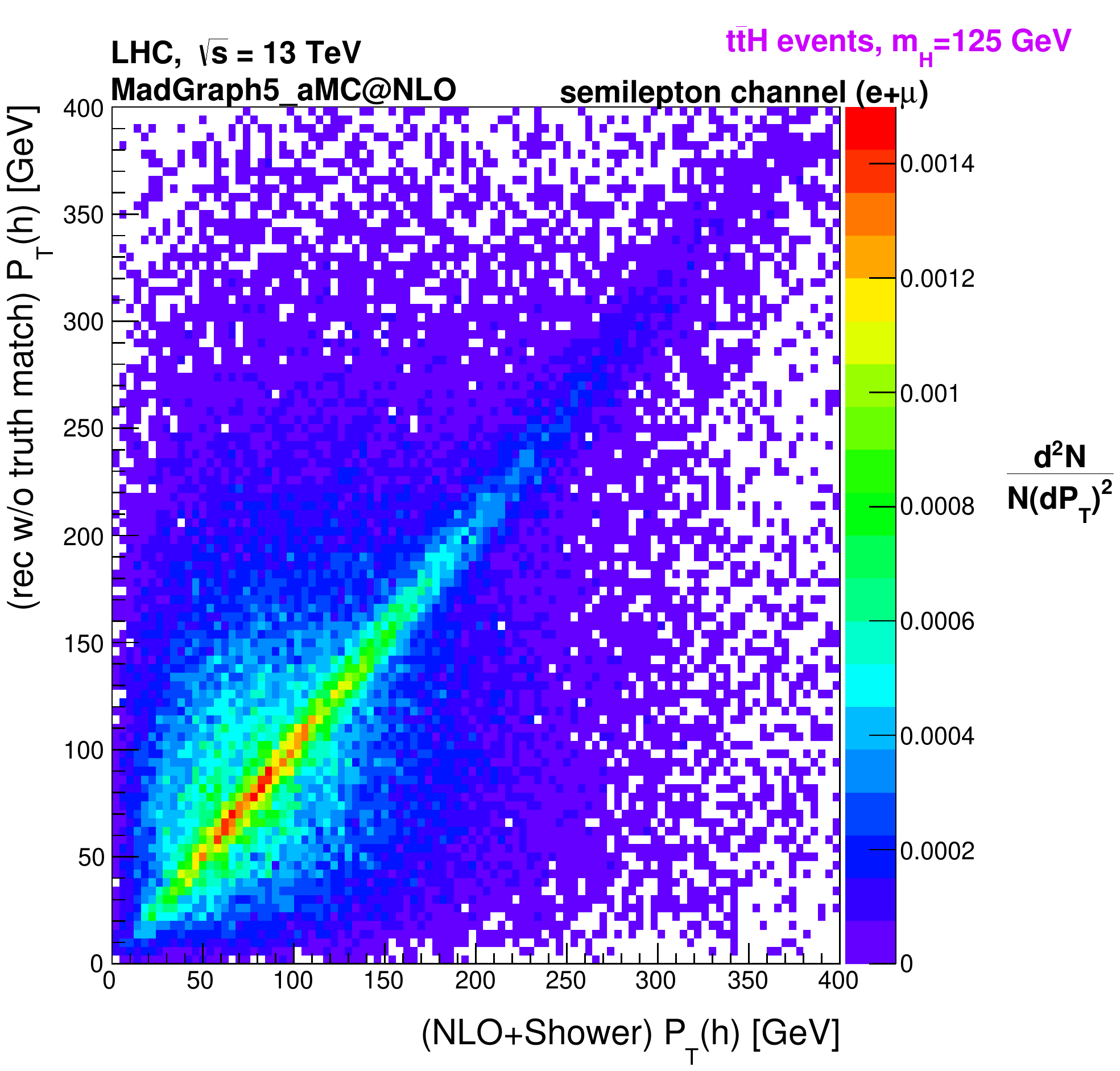,height=6.3cm,width=8cm,clip=} 		& \quad & \epsfig{file=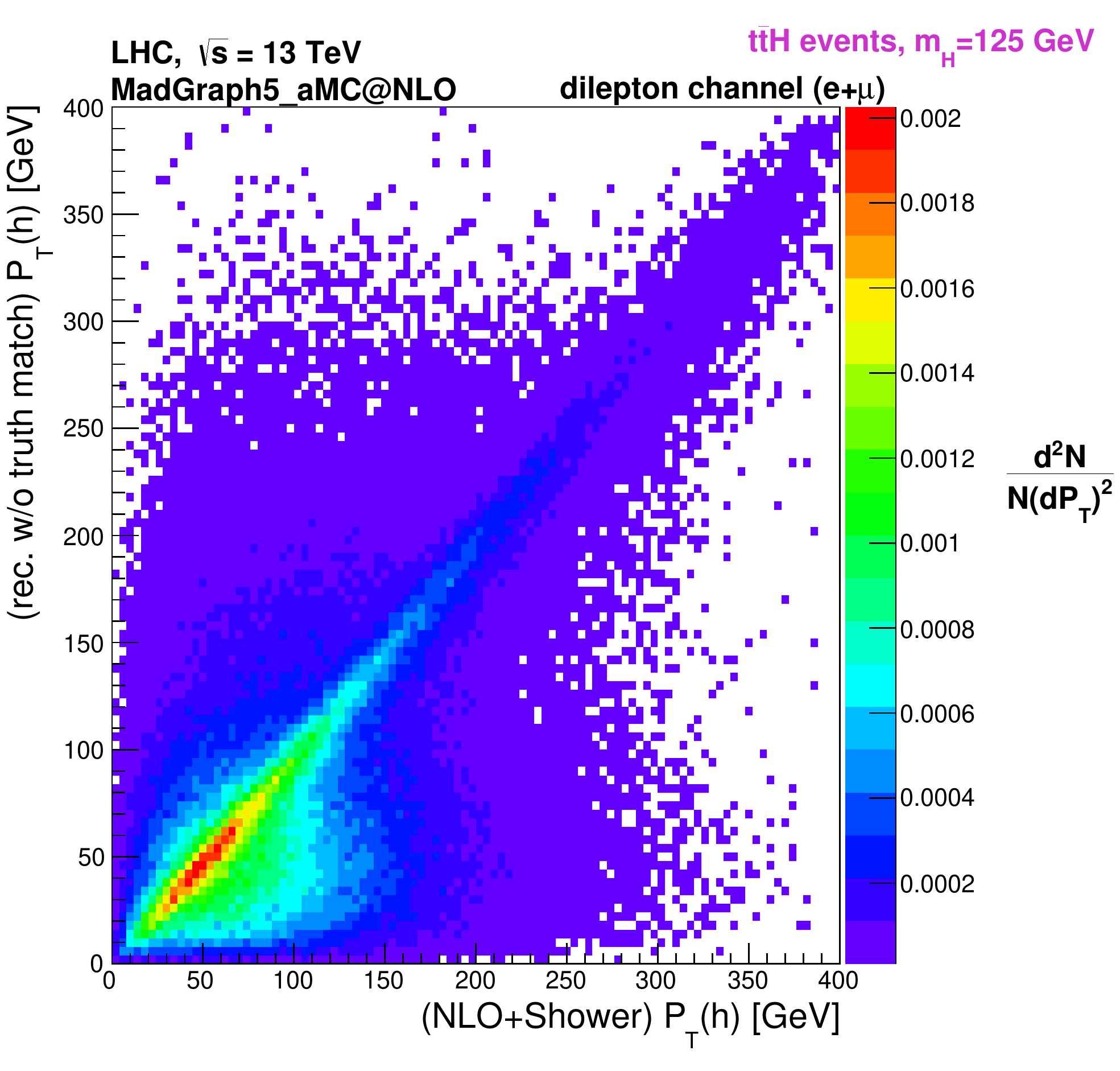,height=6.3cm,width=8cm,clip=} \\
\epsfig{file=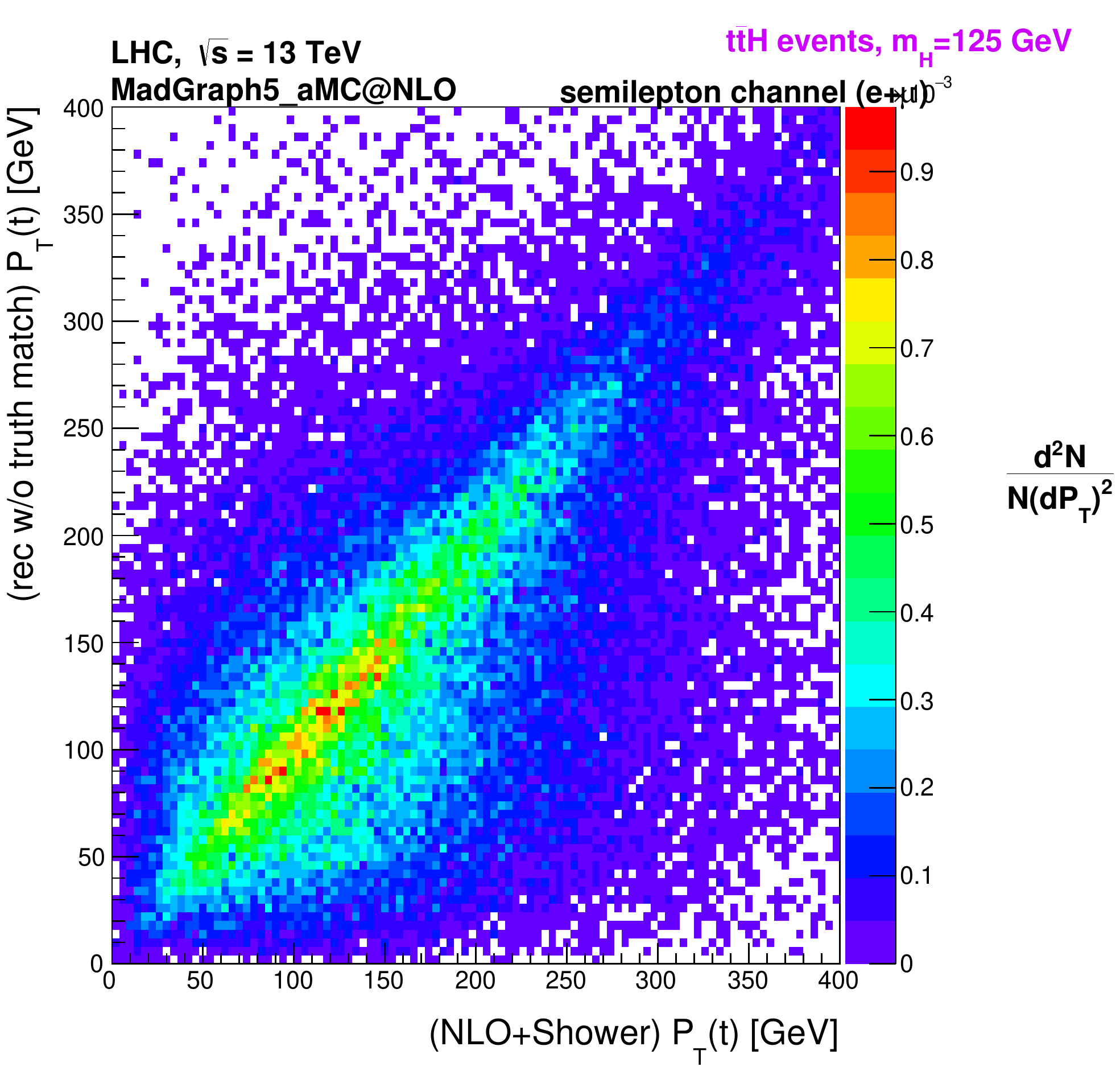,height=6.3cm,width=8cm,clip=} 		& \quad & \epsfig{file=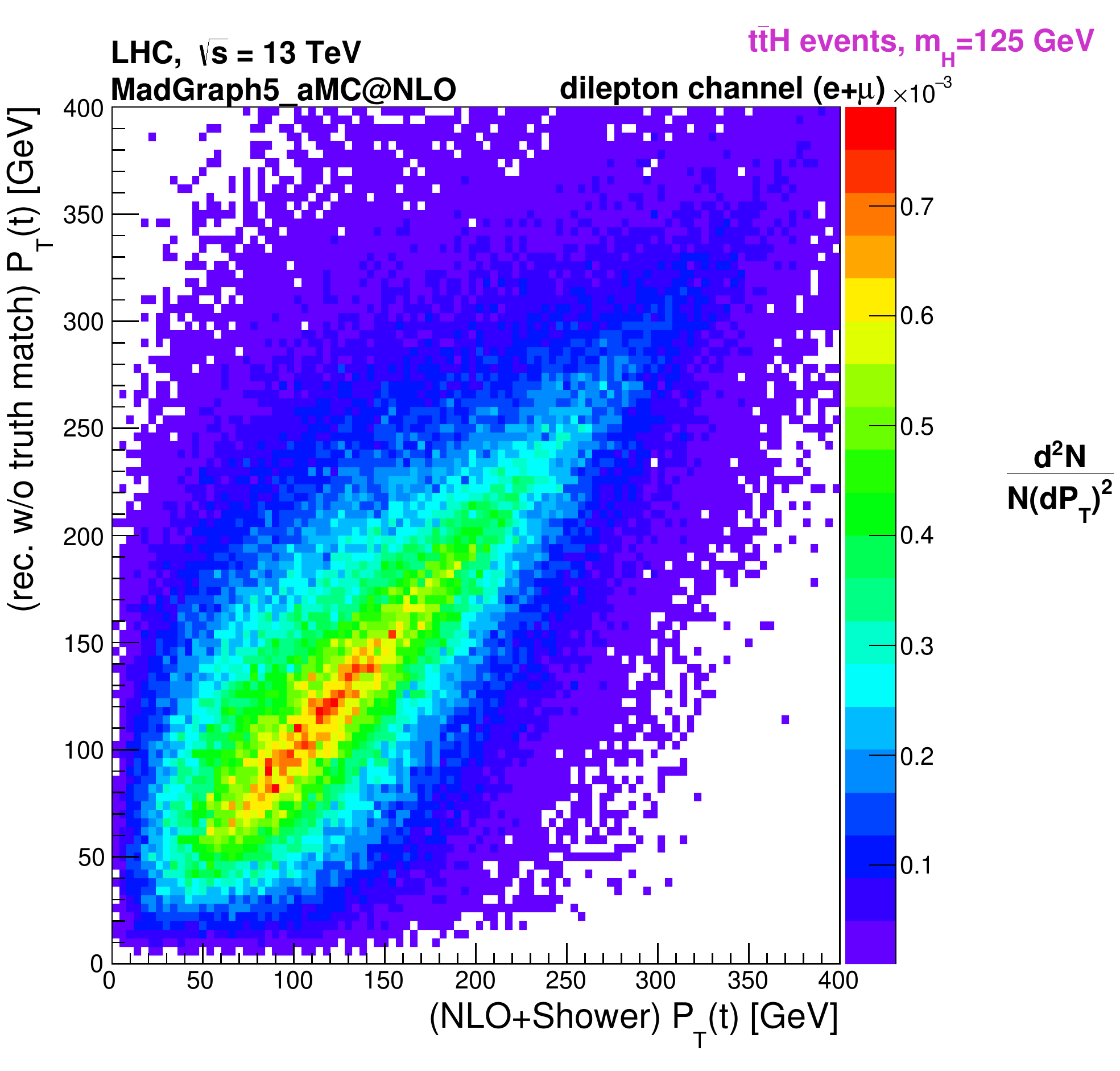,height=6.3cm,width=8cm,clip=} \\[-2mm]
\end{tabular}
\caption{Reconstructed transverse momenta of the Higgs boson (top-left) and top quark (bottom-left), as a function of the values obtained at parton level, for semileptonic decays of $t\bar th$ signal events. The corresponding distributions for dileptonic $t\bar th$ signal events, are shown on the right. The ({\it rec w/o truth match}) label, refers to final objects from the {\scshape KLFitter} full kinematic reconstruction, without trying any matching with leptons, quarks and bosons from parton level. The color bar (at the right end of each plot) represents the  total number of events in each bin with the red code on top corresponding to more events than the blue code on bottom.}
\label{fig:genexp2D}
\end{center}
\end{figure*}

In Table~\ref{tab:cutflow} the expected cross-sections (in fb) are shown, at {\it pre-selection} and {\it final selection} levels, for semileptonic final states of $t\bar th$ signals and SM backgrounds. The $t\bar{t}A$ pseudo-scalar signal was scaled to the $t\bar{t}H$ scalar cross-section for illustration purposes only. Figure~\ref{fig:genexp2D} shows the reconstructed transverse momenta of the Higgs boson (top-left) and top quark (bottom-left) against the corresponding values at generator level. Equivalent results for the $t\bar th$ dileptonic final state, published in~\cite{AmorDosSantos:2017ayi}, are presented on the right-hand-side for comparison. The reconstruction using {\scshape KLFitter} performs better  than the one used in the dileptonic channel, avoiding for instance the asymmetries seen in the transverse momentum of the Higgs boson (top-left). 
This is particularly relevant, since the shape of the $p_T(h)$ distribution is particularly sensitive to the CP nature of the Higgs boson Yukawa couplings to top quarks. Although the kinematic fit could be further improved, no optimisation of the reconstruction was attempted.

\begin{table}[h]
	\renewcommand{\arraystretch}{1.3}
	\begin{center}
		\begin{tabular}{|l|c|c|}
			\hline
			&    $\sigma$(fb)    &    $\sigma$(fb)    \\[-1mm]
			&   Pre-Selection   &   Final Selection  \\[-1mm]
			\hline  
			$t\bar{t}$+$c\bar{c},t\bar{t}$+lf    		&    	     2488    	           &       565.5          	\\         
			$t\bar{t}$+$b\bar{b}$    				&    	       898.4    	           &       165.6          	\\     
			$t\bar{t}$+$V(V$=$Z,W)$    			&    	        74.9    	           &           4.1          	\\ 
			Single $t$    						&    	       492.2      	      	  &           4.9           	\\
			$W$+jets     						&    	     3293      	           &         	0          	\\              
			$W$+$b\bar{b}$    					&    	       709.7     	           &           3.7     	\\              
			Diboson    						&             996.6      	           &           0.5         	\\         
			\hline  
			Total back.    						&    	     8953   	           	   &       744.3          	 \\             
			\hline \hline  
			$t\bar{t}H$    						&            26.6     	           &           8.85        	 \\      
			$t\bar{t}A$    						&            18.9      	       	   &           6.07         	 \\     
			\hline
		\end{tabular}
		\caption{{Expected cross-sections (in fb) at {\it pre-selection} and {\it final selection} levels, for $t\bar th (h=H,A)$ signals and SM backgrounds, at a centre-of-mass energy of 13~TeV, at the LHC.}}
		\label{tab:cutflow}
	\end{center}
\end{table}

\section{III. Angular Observables}

\begin{figure*}
	\vspace*{5cm}
\begin{center}
\begin{tabular}{ccc}
\epsfig{file=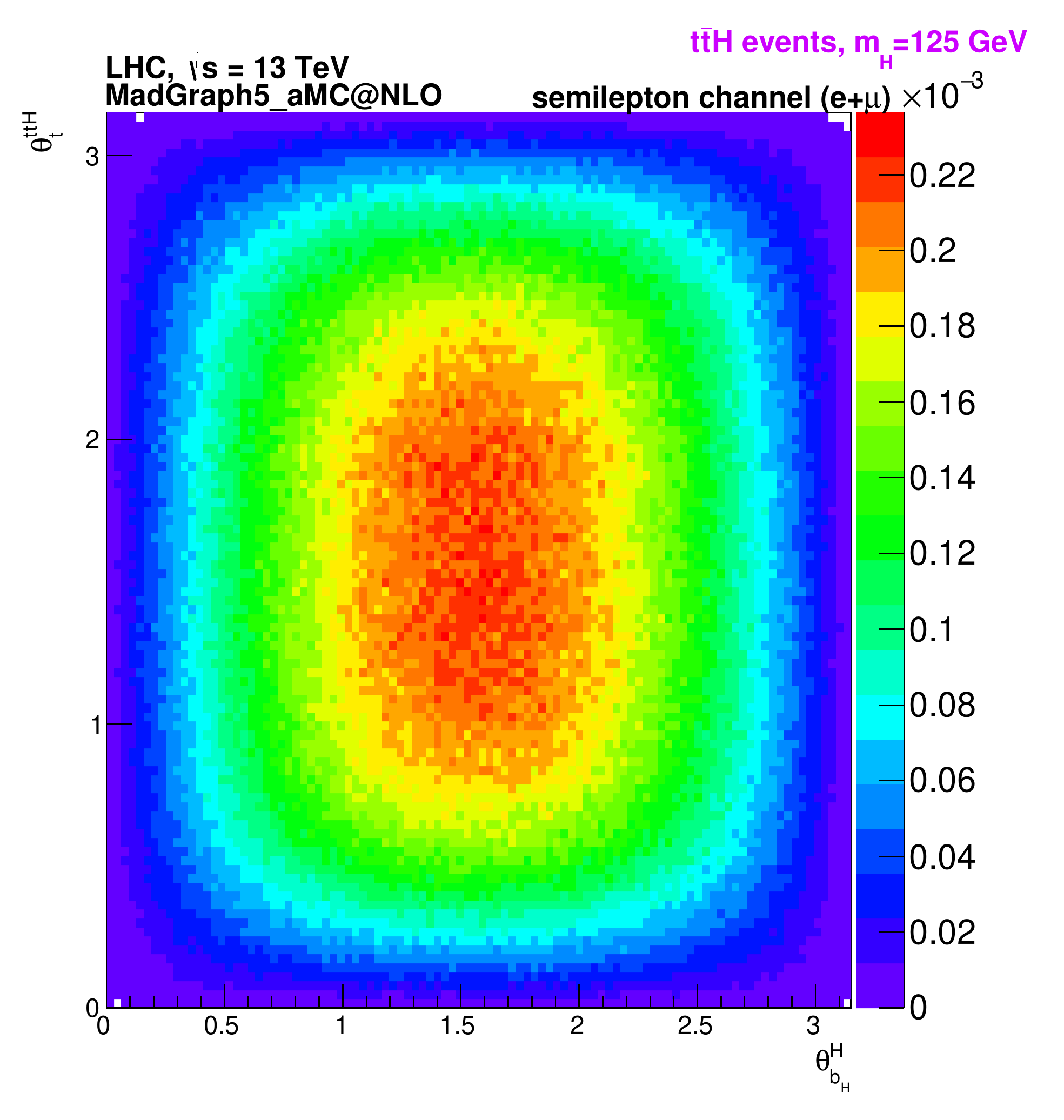,height=7.5cm,width=7.5cm,clip=} 		& \quad & \epsfig{file=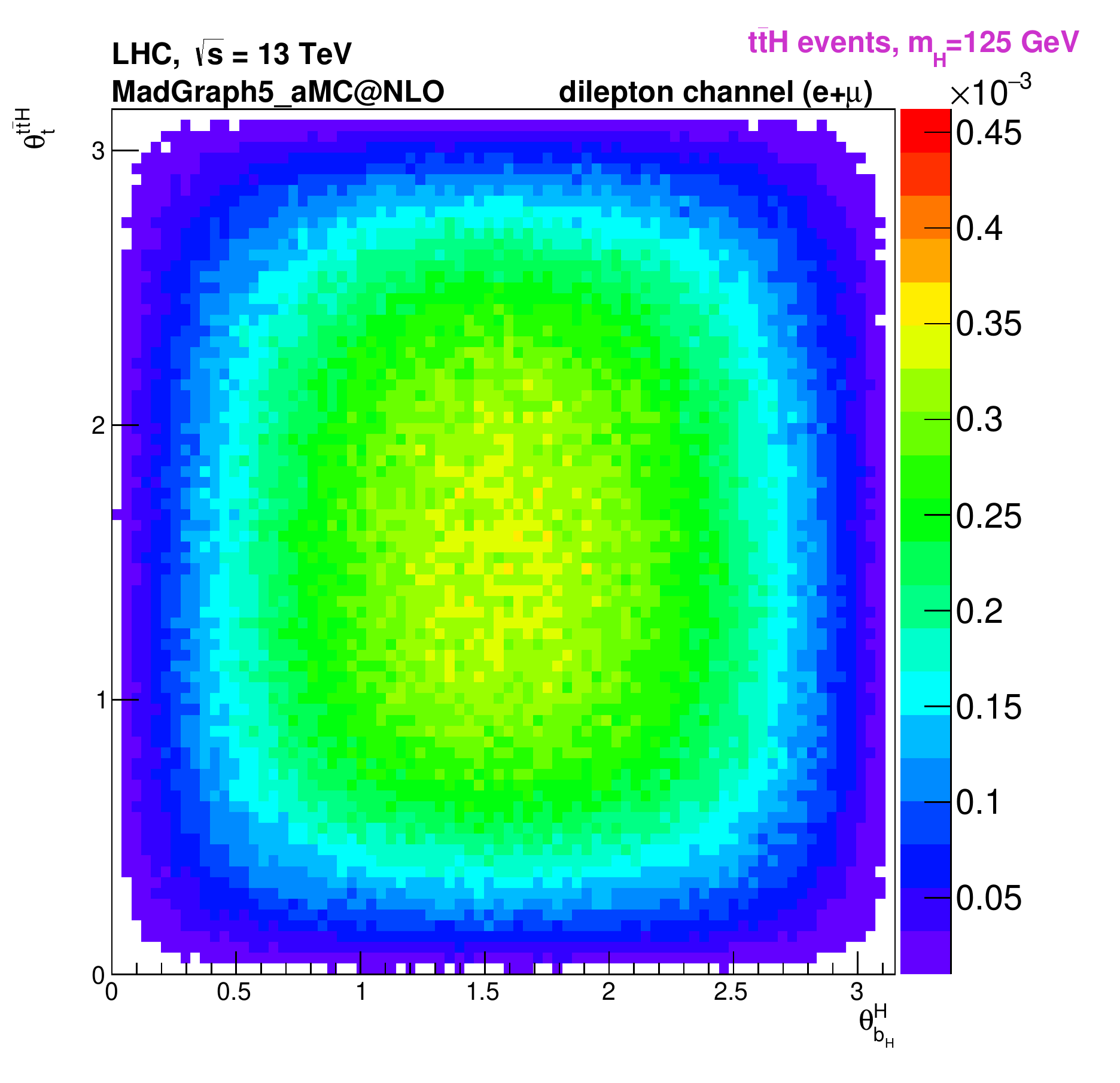,height=7.5cm,width=7.5cm,clip=}  \\
\epsfig{file=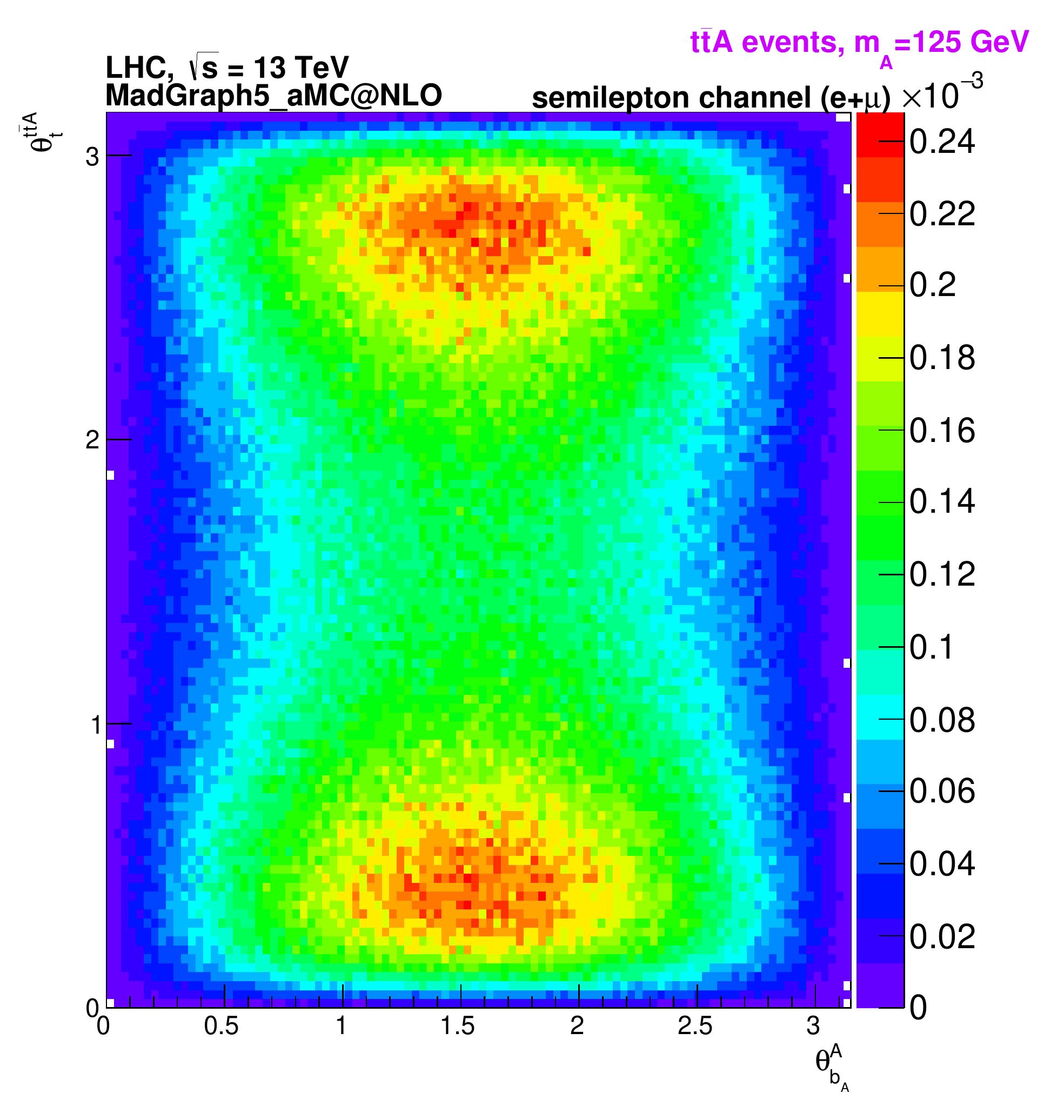,height=7.5cm,width=7.5cm,clip=}  		& \quad & \epsfig{file=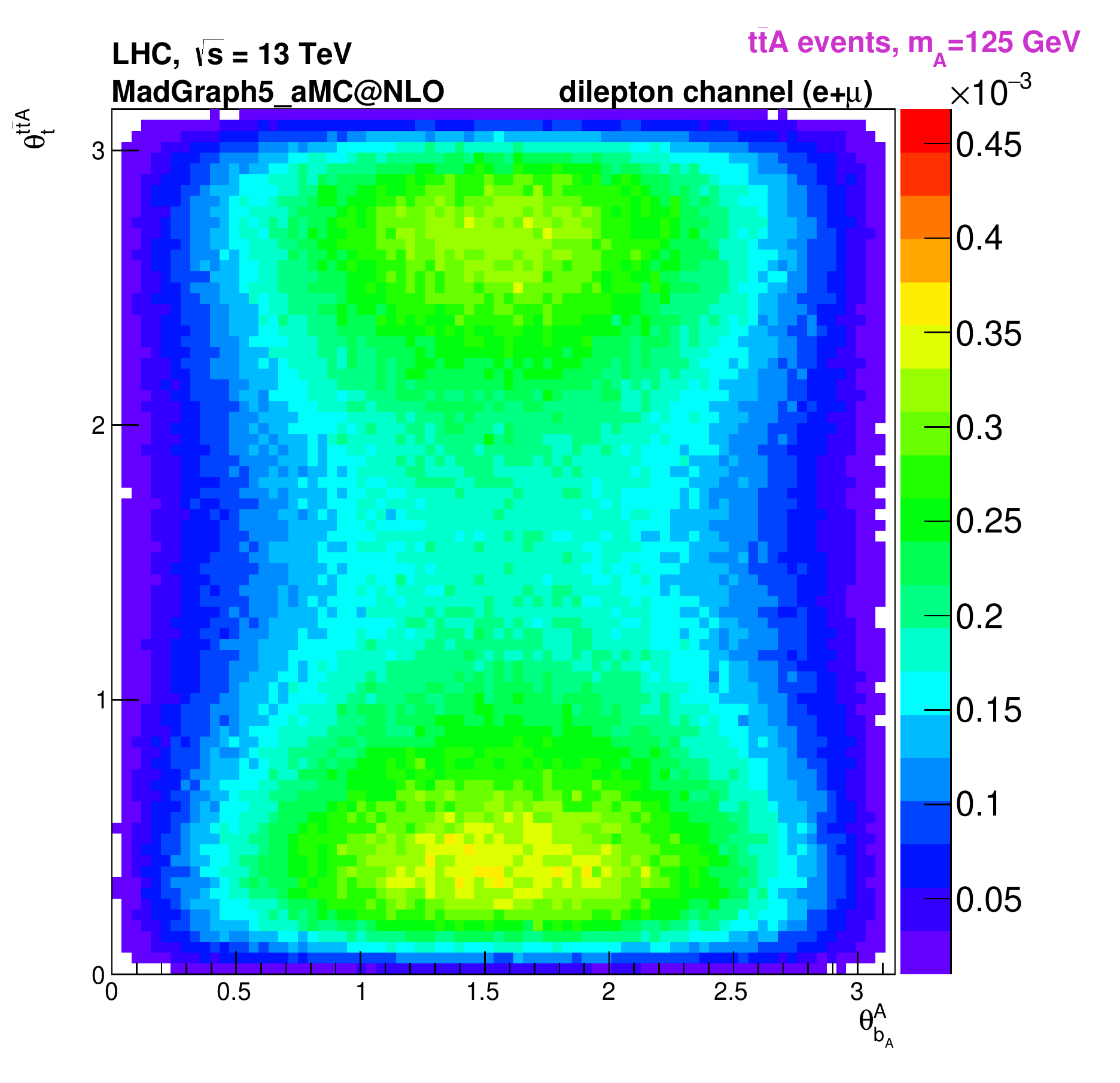,height=7.5cm,width=7.5cm,clip=}\\
\end{tabular}
\caption{The angle between the momentum direction of the top quark, in the $t\bar th$ system, and the $t\bar th$ direction, in the lab frame ($\theta^{t\bar th}_t$), as a function of the angle  
between the Higgs momentum, in the $\bar th$ frame, and the momentum of the $b$ quark from the Higgs boson, in the Higgs boson frame ($\theta^h_{b_h}$). 
The top (bottom) distributions are for the scalar $h=H$ (pseudo-scalar, $h=A$) $t\bar th$ signal. The plots on the left (right) show the semileptonic (dileptonic) channel of $t\bar th$ decays.}
\label{fig:ttHangles01}
\end{center}
\end{figure*}

\begin{figure*} 
\vspace*{5cm}
\begin{center}
\begin{tabular}{ccc}
	\epsfig{file=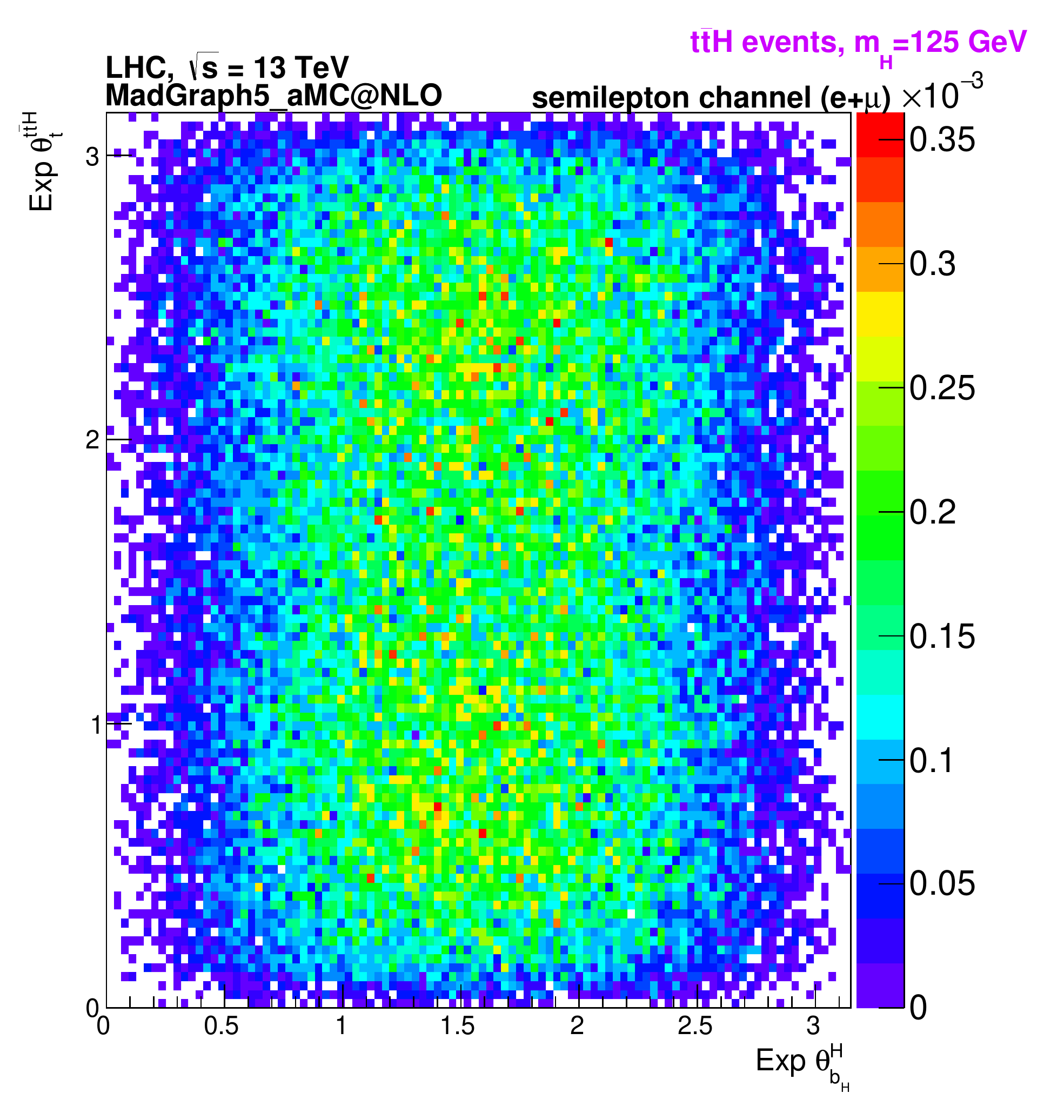,height=7.5cm,width=7.5cm,clip=} 	& \qquad & \epsfig{file=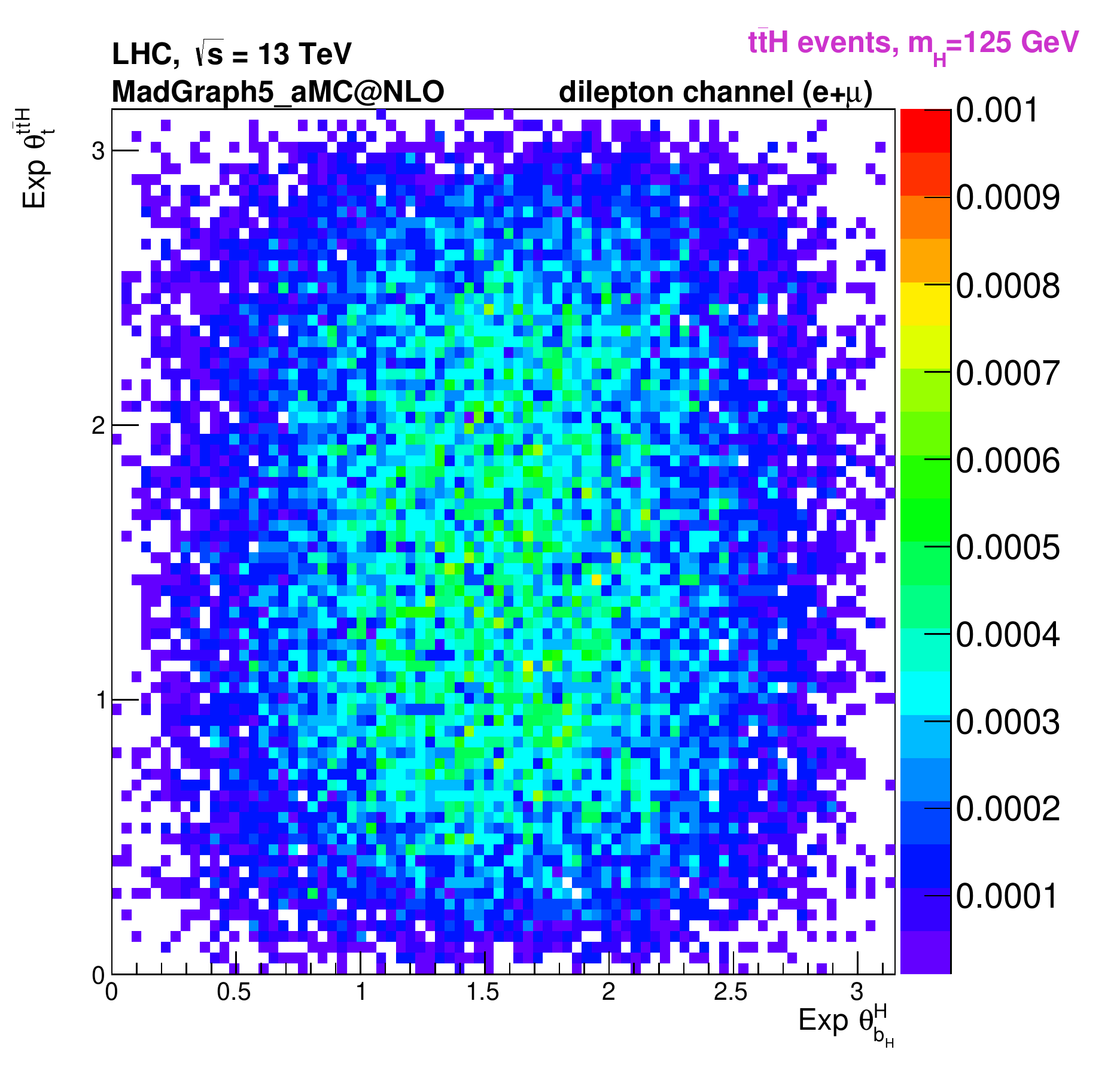,height=7.5cm,width=7.5cm,clip=} \\
	\epsfig{file=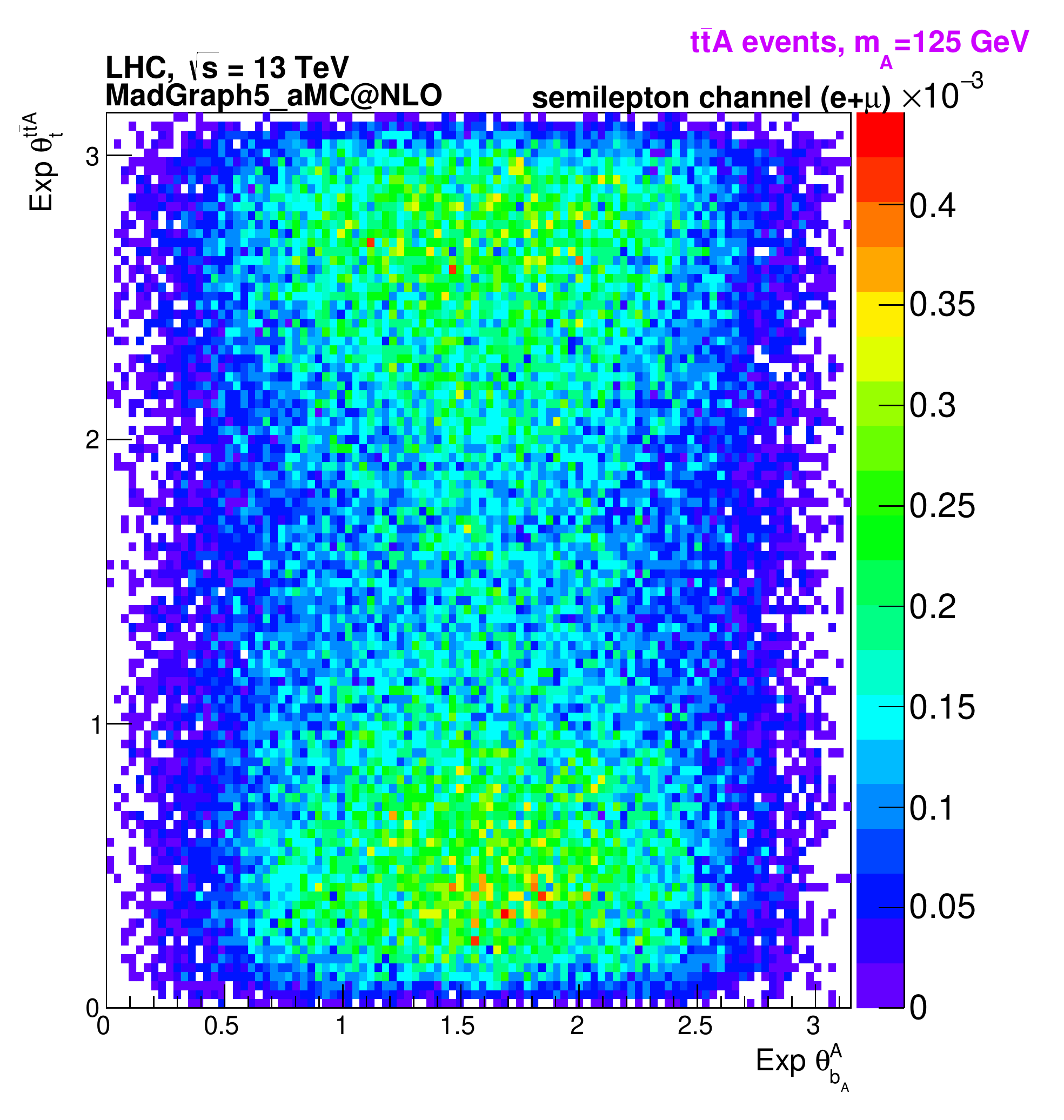,height=7.5cm,width=7.5cm,clip=}  	& \qquad & \epsfig{file=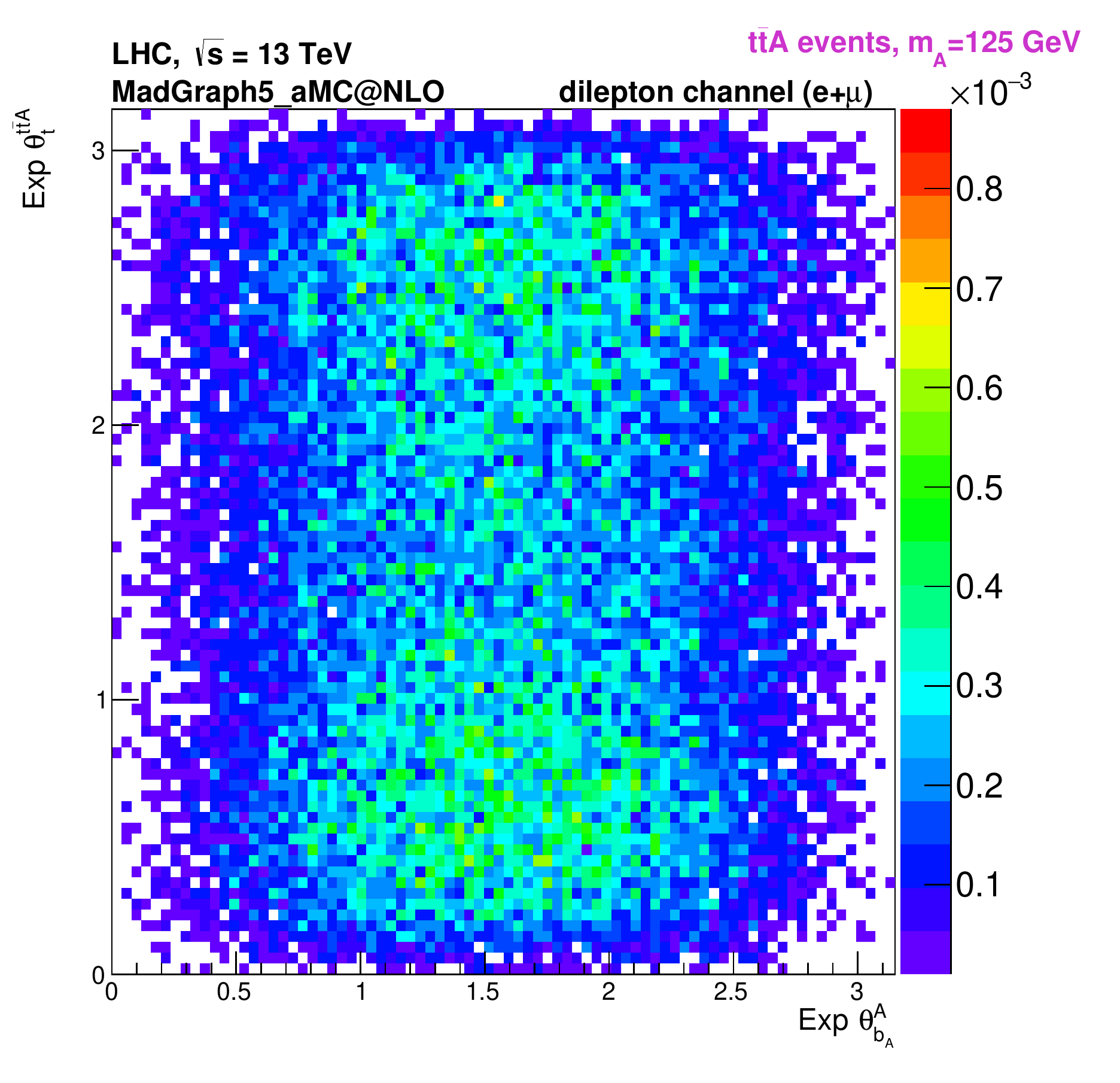,height=7.5cm,width=7.5cm,clip=}\\
\end{tabular}
\caption{The same as Figure~\ref{fig:ttHangles01}, after event selection and full kinematic reconstruction.}
\label{fig:ttHangles02}
\end{center}
\end{figure*}

For the reconstruction of the angular distributions we use the spin helicity formalism and, generically, define \(\theta_Y^X\) as the angle between the  momentum direction of the $Y$ particle (or system), measured in the rest frame of $X$, with respect to the direction of $X$, in the rest frame of its parent particle~\cite{Santos:2015dja}. As particles follow successive decays starting from the \(t\bar{t}h\) centre-of-mass system ($X=t\bar th$) until all intermediate particles have decayed, $X$ defined above include 3-, 2- and single-particle systems. The $t\bar th$ system momentum direction is measured with respect to the laboratory frame. In case of ambiguity in describing the angular distributions, the exact definition for the angles is specified in the text. In performing the boosts, two different prescriptions can be used for the decays: (1) the {\it direct} approach, when the laboratory four-momentum of particles were used for $X$ and $Y$, or (2) the {\it sequential} approach, where  the four-momentum of particles $X$ and $Y$ were boosted through all intermediate centre-of-mass systems. Both prescriptions lead to different distributions due to the non-Abelian nature of the Lorentz group.
In Figure~\ref{fig:ttHangles01} and Figure~\ref{fig:ttHangles02}, we show \(\theta^{t\bar{t}h}_t\) (the angle between the momentum direction of the top quark, in the \(t\bar{t}h\) system, and the \(t\bar{t}h\) direction, in the lab frame) versus \(\theta^{h}_{b_h}\) (the angle between the momentum of the $b$ quark from the Higgs boson, in the Higgs boson frame, and the Higgs boson momentum in the $\bar t h$ frame). Distributions are shown at generator level in Figure~\ref{fig:ttHangles01} and after event selection and kinematic reconstruction in Figure~\ref{fig:ttHangles02}. While the left (right) distributions are for  $t\bar th$ semileptonic (dileptonic) decays, the top (bottom) ones are for the scalar $h=H$ (pseudo-scalar $h=A$) $t\bar t h$ signal. The dileptonic results are only shown for comparison. 
The pattern differences observed between the scalar and pseudo-scalar signal distributions are quite noticeable, even after event selection and kinematic reconstruction. 
This behaviour is particularly visible in $t\bar th$ semileptonic decays after kinematic reconstruction (Figure~\ref{fig:ttHangles02}). 
While events tend to be more uniformly distributed in the plot for the case of the scalar couplings (Figure~\ref{fig:ttHangles02}-top), the pseudo-scalar case tends to concentrate events in two extreme regions (Figure~\ref{fig:ttHangles02}-bottom).

Given the good performance of the kinematic reconstruction in semileptonic decays of $t\bar th$, we study the following angular distributions and corresponding asymmetries defined in~\cite{Santos:2015dja}:

\begin{tabular}{ccl}
	& \quad & $\cos{(\theta^{\bar{t}h}_{h})}\cos{(\theta^{h}_{\ell^-})}$ and $A^{\ell-(h)}_{FB}$(\textit{direct} boost), \\[1mm]   
	& \quad & $\sin{(\theta^{t\bar{t}h}_{h})}\sin{(\theta^{\bar{t}}_{\bar{b}_{\bar{t}}})}$ and $A^{\bar{b}_{\bar{t}}({\bar{t}})}_{FB}$({\it sequential}~boost), \\[1mm]
	& \quad & $\sin{(\theta^{t\bar{t}h}_{h})}\cos{(\theta^{\bar{t}}_{b_h})}$ and $A^{b_h(\bar{t})}_{FB}$({\it sequential}~boost), \\[1mm]
	& \quad & $\sin{(\theta^{t\bar{t}h}_{t})}\sin{(\theta^{h}_{W+})}$ and $A^{W+({h})}_{FB}$({\it sequential}~boost), \\[1mm]
	& \quad & $\sin{(\theta^{t\bar{t}h}_{\bar{t}})}\sin{(\theta^{h}_{b_h})}$ and $A^{b_h(h)}_{FB}$({\it sequential}~boost), \\[1mm]     
	& \quad & $\sin{(\theta^{t\bar{t}h}_{h})}\sin{(\theta^{t\bar{t}}_{\bar{t}})}$ and $A^{\bar{t}(t\bar{t})}_{FB}$(\textit{direct} boost) and \\[1mm]
	& \quad & $b_4 = (p^z_t . p^z_{\bar{t}}) / (|\vec{p}_{t}| . |\vec{p}_{\bar{t}}| )$ and $A^{b_4}_{FB}$, as defined in~\cite{Gunion:1996xu}. \\[3mm]   
\end{tabular}
 
Table~\ref{tab:AsymGenExp} shows the asymmetry values for the different scalar and pseudo-scalar $t\bar th$ signals, together with the ones expected for the dominant SM background, $t\bar t b \bar b$. These were calculated after event selection and full kinematic reconstruction. 

\begin{table}[h]
	\renewcommand{\arraystretch}{1.2}
	\begin{center}
		\begin{tabular}{|c|c|c|cc|}
			\hline
						               					& 		\multicolumn{3}{c}{Final Selection }                &\\
		~~~Asymmetries~~~							&		\multicolumn{3}{c}{and Kinematic Reconstruction} 	        &\\
												& 			$t\bar{t}H$ 	      	& 	$t\bar{t}A$       & $t\bar{t}b\bar{b}$ 	&\\
			\hline
			$A^{\ell-(h)}_{FB}$    					&   	  ~~~$+0.10 $~~~	& 	  ~~~$+0.17 $~~~	&   ~~$-0.01 $~~ &\\               
			$A^{\bar{b}_{\bar{t}}({\bar{t}})}_{FB}$			&   		$+0.20 $		&		$+0.19 $		&       $-0.09 $ &\\
			$A^{b_h({\bar{t}})}_{FB}$ 					&   		$-0.67  $	 	&		$-0.72  $		&       $-0.65 $  &\\       
			$A^{W+({h})}_{FB}$						&   		$-0.33  $	 	&		$-0.51  $		&       $-0.51 $  &\\
			$A^{b_h(h)}_{FB}$    					&   		$+0.18 $	 	&		$+0.02 $		&       $-0.05 $  &\\
			$A^{\bar{t}(t\bar{t})}_{FB}$    				&   		$+0.17 $	 	&		$+0.15 $		&       $-0.11 $  &\\
			$A^{b_4}_{FB}$    						&   		$+0.17 $	 	&		$-0.08  $		&       $+0.06 $  &\\   
			\hline
		\end{tabular}
		\caption{{Asymmetry values for $t\bar{t}H$, $t\bar{t}A$ and $t\bar{t}b\bar{b}$ after selection criteria and kinematic reconstruction, are shown, for semileptonic final states of the $t\bar{t}$ system.} 
		The Monte Carlo statistical uncertainties of all asymmetries were evaluated to be below 10$^{-2}$.}
		\label{tab:AsymGenExp}
	\end{center}
\end{table}

%\begin{table}[h]
%	\renewcommand{\arraystretch}{1.2}
%	\begin{center}
%		\begin{tabular}{|c|c|cc|}
%			\hline
%						               					& 		\multicolumn{2}{c}{Final Selection}                &\\
%		Asymmetries					&		\multicolumn{2}{c}{and Kinematic Rec.} 	        &\\
%												& 		$t\bar{t}H$~/~$t\bar{t}A$       & $t\bar{t}b\bar{b}$ 	&\\
%			\hline
%			$A^{\ell-(h)}_{FB}$    					&   	~~$+0.104\pm0.005$~/~$+0.166\pm0.005$~~		& ~$-0.005\pm0.007$~  &\\               
%			$A^{\bar{b}_{\bar{t}}({\bar{t}})}_{FB}$			&   	~~$+0.204\pm 0.004$~/~$+0.186\pm0.004$~~		& ~$-0.086\pm0.005$~  &\\
%			$A^{b_h({\bar{t}})}_{FB}$ 					&   	~~$-0.672\pm0.003$~/~$-0.717\pm0.003$~~		& ~$-0.650\pm0.004$~  &\\       
%			$A^{W+({h})}_{FB}$						&   	~~$-0.327\pm0.004$~/~$-0.512\pm0.003$~~		& ~$-0.513\pm0.004$~  &\\
%			$A^{b_h(h)}_{FB}$    					&   	~~$+0.184\pm0.004$~/~$+0.015\pm0.004$~~		& ~$-0.050\pm0.005$~  &\\
%			$A^{\bar{t}(t\bar{t})}_{FB}$    				&   	~~$+0.156\pm0.004$~/~$+0.154\pm0.004$~~		& ~$-0.107\pm0.005$~  &\\
%			$A^{b_4}_{FB}$    						&   	~~$+0.167\pm0.004$~/~$-0.076\pm0.004$~~		& ~$+0.058\pm0.005$~  &\\   
%			\hline
%		\end{tabular}
%		\caption{{Asymmetry values for $t\bar{t}H$, $t\bar{t}A$ and $t\bar{t}b\bar{b}$ after selection criteria and kinematic reconstruction, are shown.}}
%		\label{tab:AsymGenExp}
%	\end{center}
%\end{table}

\begin{figure*} 
\vspace*{3cm}
\begin{center}
\begin{tabular}{cc}	
\hspace*{-5mm} \epsfig{file=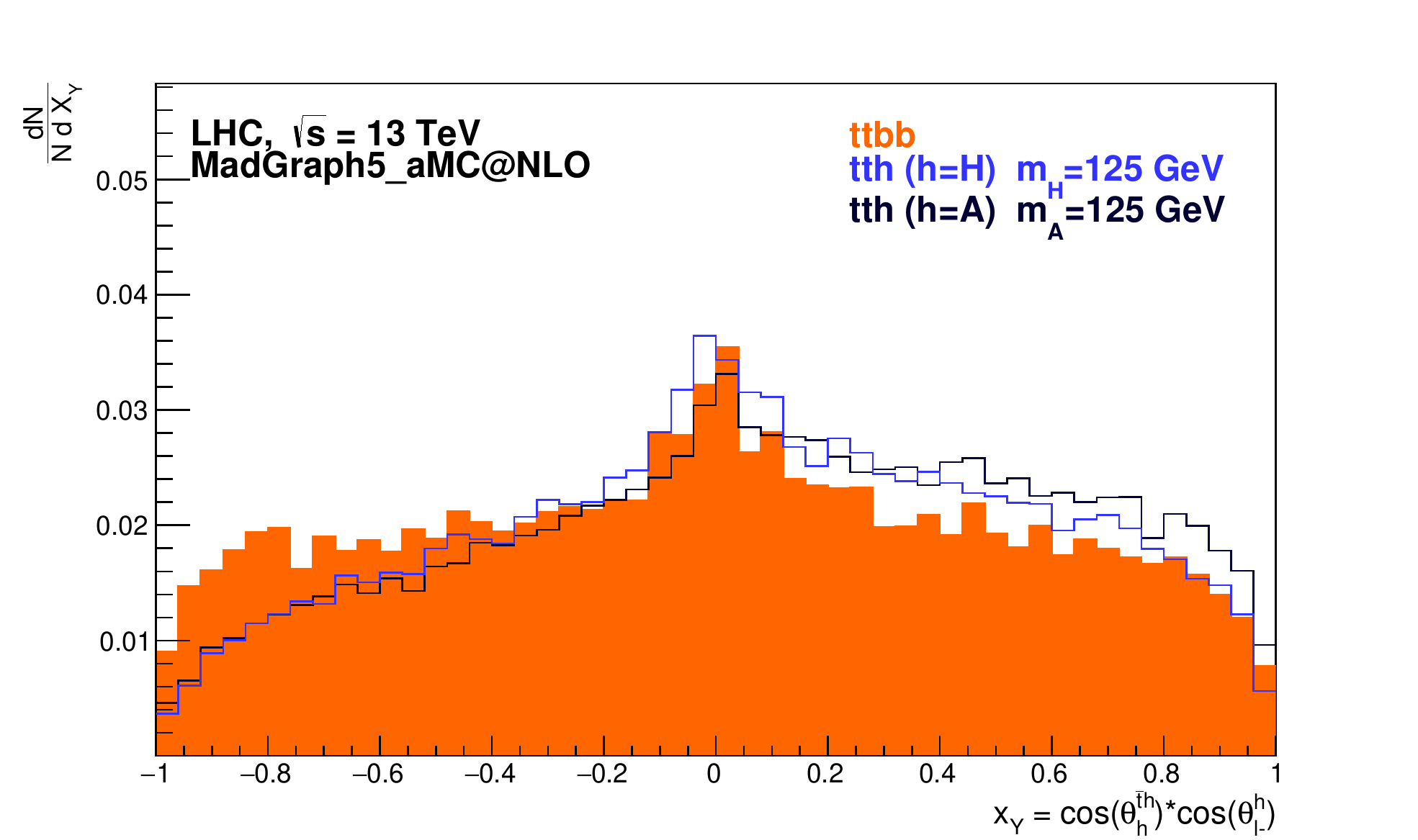,height=5.5cm,clip=} 	& \epsfig{file=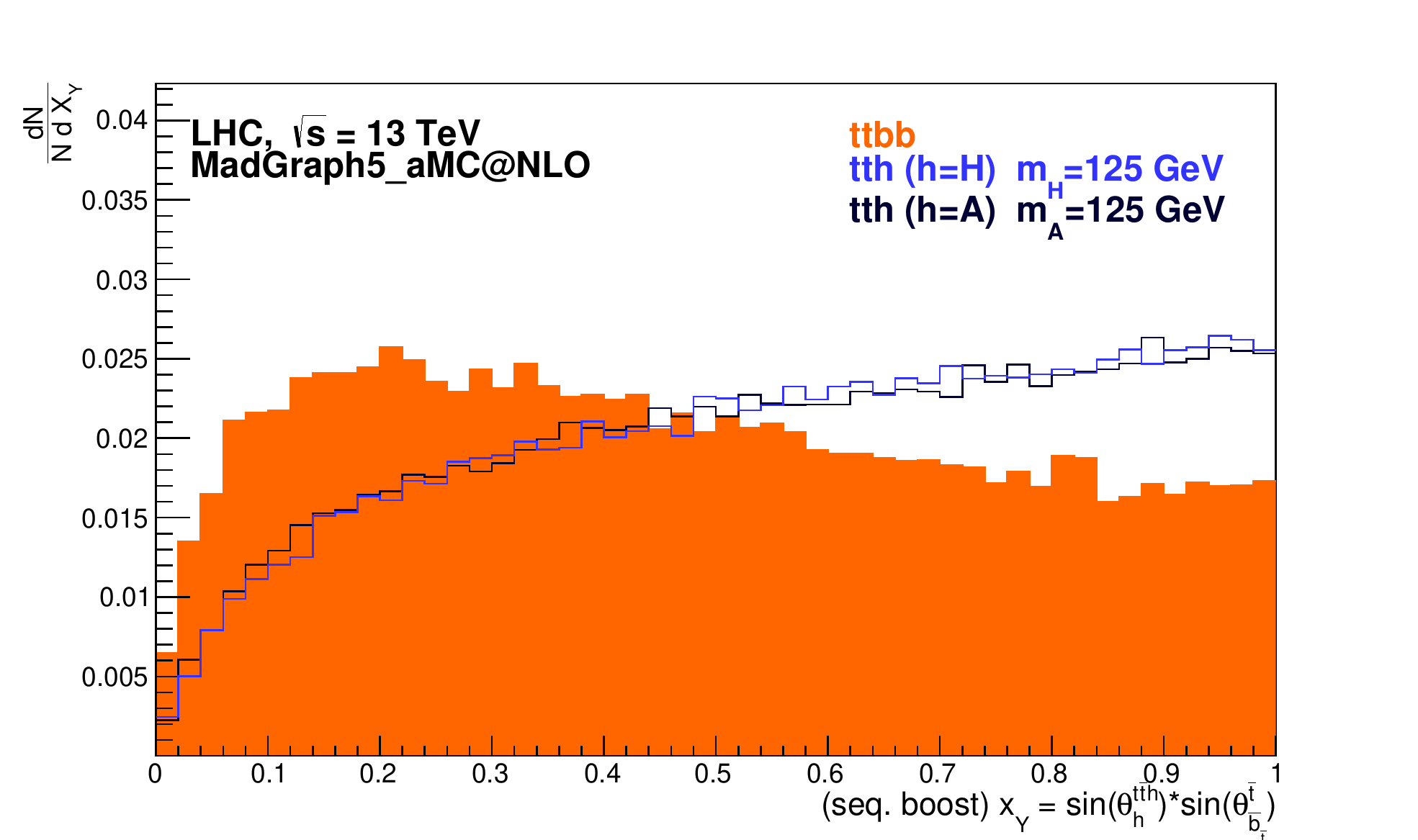,height=5.5cm,clip=}\\
\hspace*{-5mm} \epsfig{file=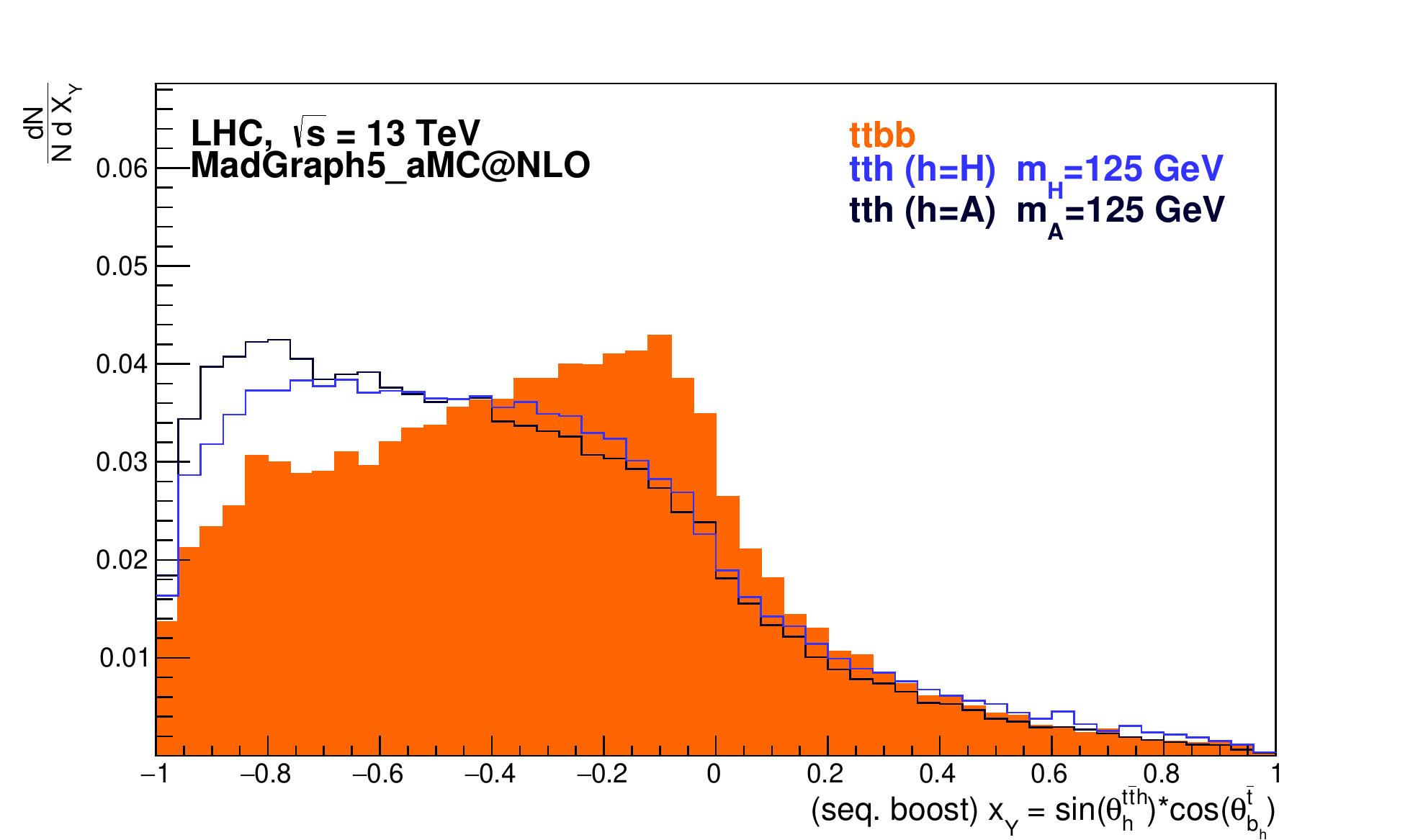,height=5.5cm,clip=} 	& \epsfig{file=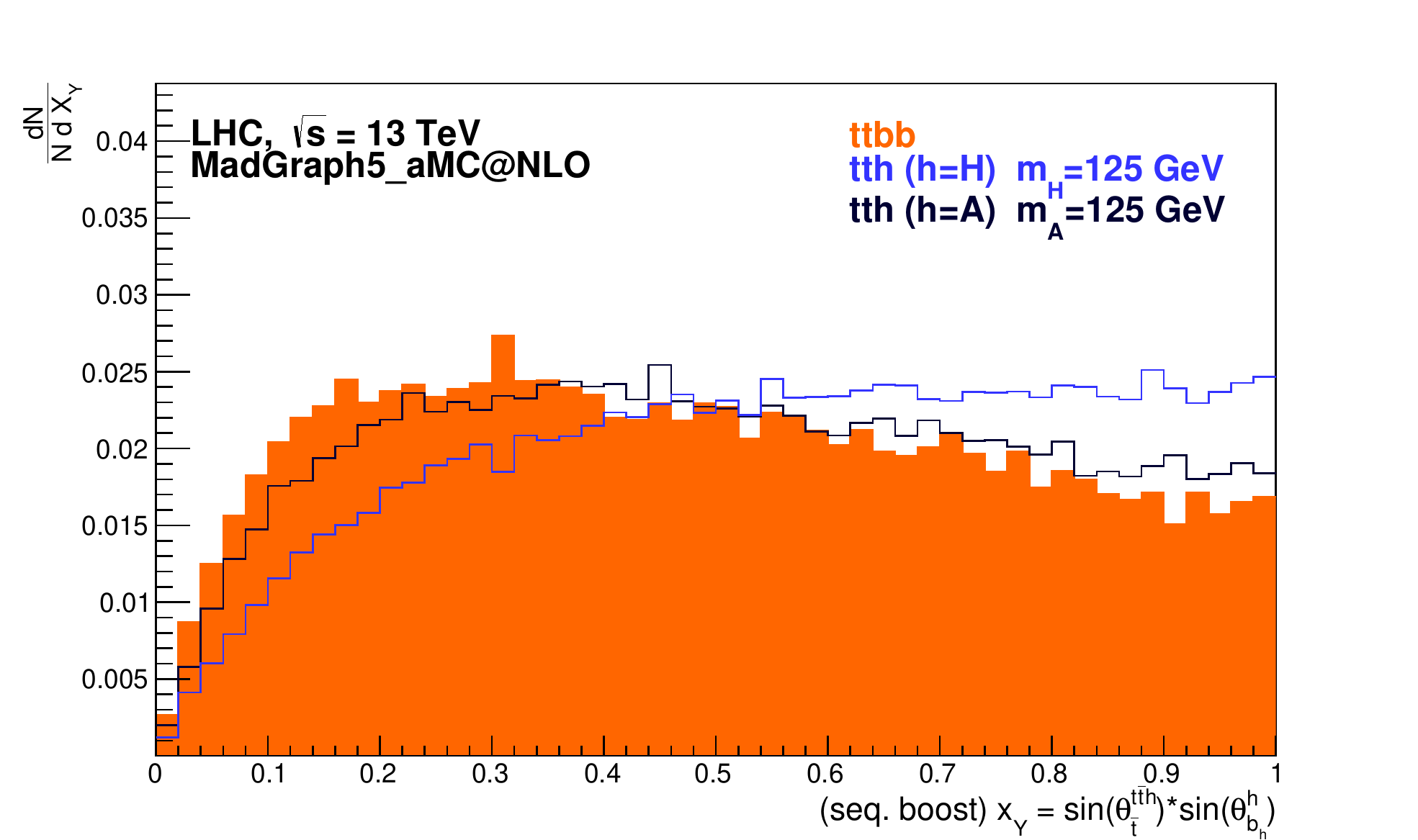,height=5.5cm,clip=}\\
\hspace*{-5mm} \epsfig{file=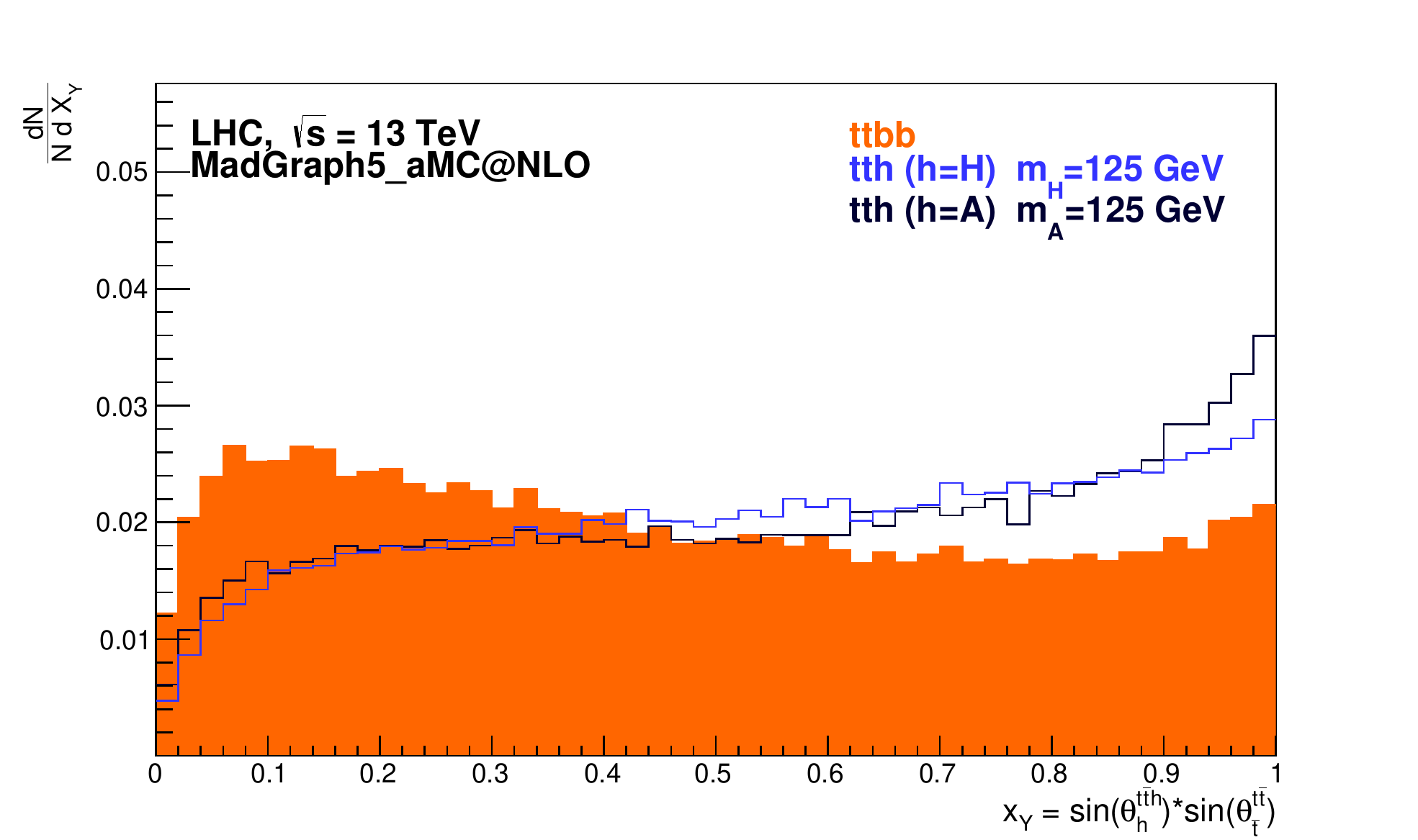,height=5.5cm,clip=} 	& \epsfig{file=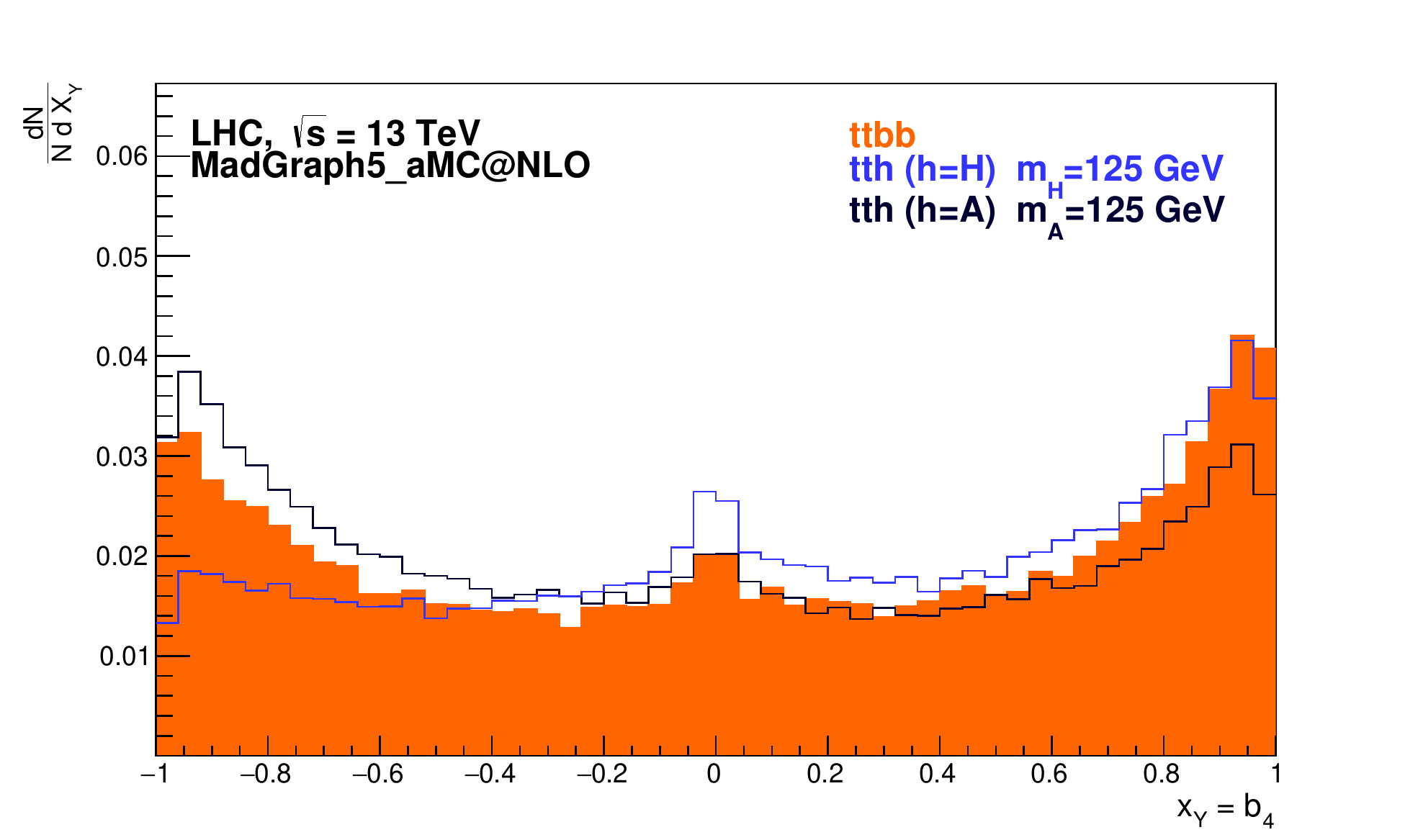,height=5.5cm,clip=}\\[-4mm]
\end{tabular}
\caption{Angular distributions:
(top-left) $\cos{(\theta^{\bar{t}h}_{h})}\cos{(\theta^{h}_{\ell^-})}$ and 
(top-right) $\sin{(\theta^{t\bar{t}h}_{h})}\sin{(\theta^{\bar{t}}_{\bar{b}_{\bar{t}}})}$;
(middle-left) $\sin{(\theta^{t\bar{t}h}_{h})}\cos{(\theta^{\bar{t}}_{b_h})}$  and
(middle-right) $\sin{(\theta^{t\bar{t}h}_{\bar{t}})}\sin{(\theta^{h}_{b_h})}$;
(bottom-left) $\sin{(\theta^{t\bar{t}h}_{h})}\sin{(\theta^{t\bar{t}}_{\bar{t}})}$ and  
(bottom-right) $b_4$~\cite{Gunion:1996xu}. These are shown after event selection and full kinematic reconstruction. 
The light blue line represents the $t\bar{t}h$ SM model signal ($h=H$ and $CP=+1$) and the dark blue line corresponds to the pure pseudo-scalar distribution $t\bar{t}h$ ($h=A$ and $CP=-1$). 
The filled region corresponds to the $t\bar{t}b\bar{b}$ dominant background. }
\label{fig:NewAng01}
\end{center}
\end{figure*}

In Figure~\ref{fig:NewAng01}, some of the corresponding angular distributions are shown. While the {\it direct} prescription was applied to boost the lepton ($\ell^-$) to the Higgs boson ($h$) system in the top-left distribution,
$\cos{(\theta^{\bar{t}h}_{h})}\cos{(\theta^{h}_{\ell^-})}$, the {\it sequential} prescription was used in the top-right distribution, 
$\sin{(\theta^{t\bar{t}h}_{h})}\sin{(\theta^{\bar{t}}_{\bar{b}_{\bar{t}}})}$, to boost the $\bar b$ to its parent top quark system ($\bar t$). For both middle plots,
$\sin{(\theta^{t\bar{t}h}_{h})}\cos{(\theta^{\bar{t}}_{b_h})}$ and
$\sin{(\theta^{t\bar{t}h}_{\bar{t}})}\sin{(\theta^{h}_{b_h})}$ on the left and right, respectively, the {\it sequential} prescription was used to boost the $b$-quark from the Higgs boson decay, to the $\bar t$  and $h$ centre-of-mass systems, respectively.
Finally, in the bottom plots, the angular distributions of 
$\sin{(\theta^{t\bar{t}h}_{h})}\sin{(\theta^{t\bar{t}}_{\bar{t}})}$ and 
$b_4$~\cite{Gunion:1996xu}, 
are shown on the left and right, respectively. Clear differences among the shapes of both $t\bar th$ signals and also with respect to the dominant background, $t\bar t b\bar b$, are visible even after event selection and full kinematic reconstruction. 

\begin{figure*} 
\begin{center}
\begin{tabular}{ccc}	
\hspace*{-9mm}\epsfig{file=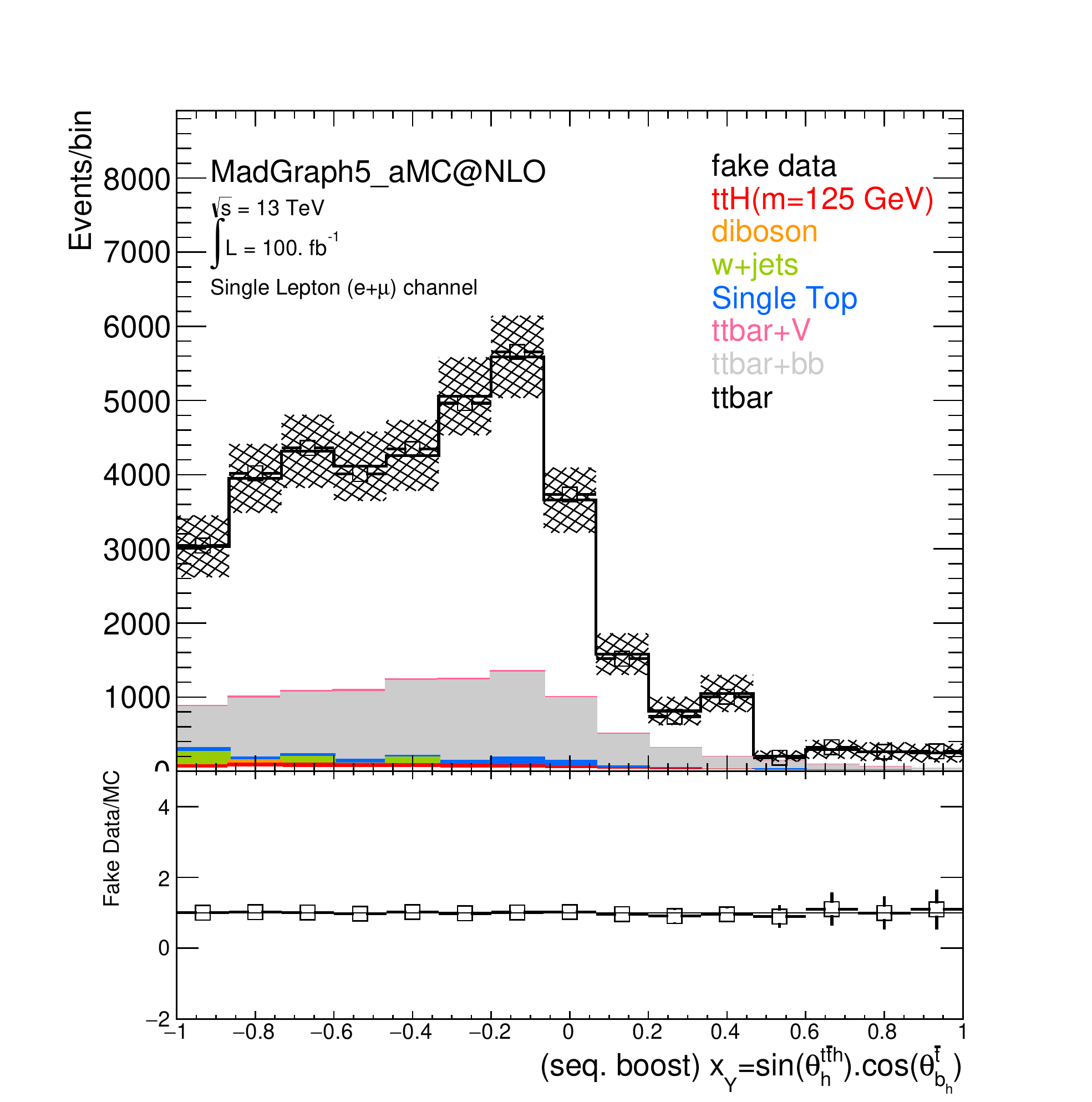,height=9cm,clip=} 		& \quad &  \hspace*{-5mm}\epsfig{file=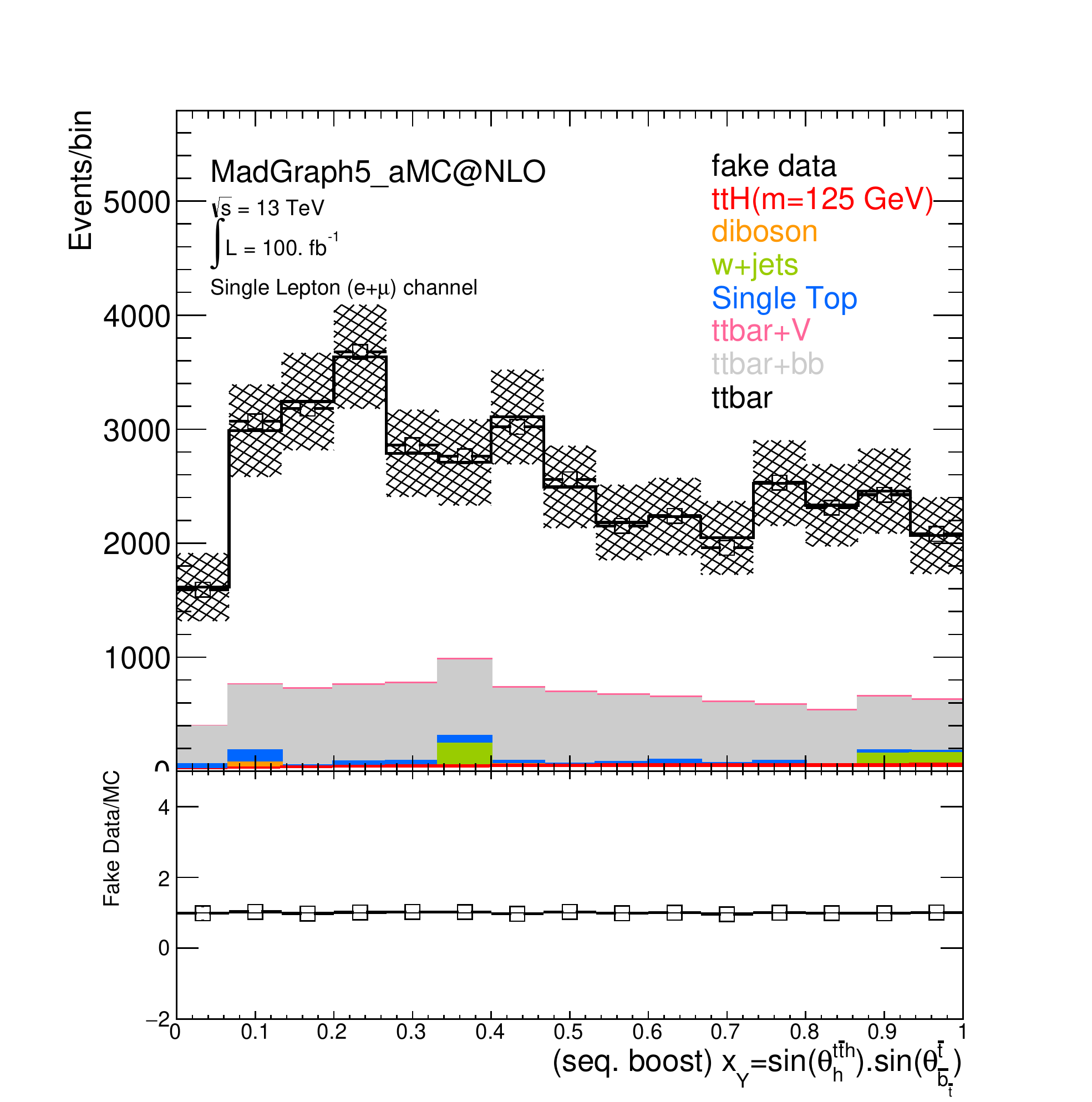,height=9cm,clip=}\\ 
\hspace*{-5mm}\epsfig{file=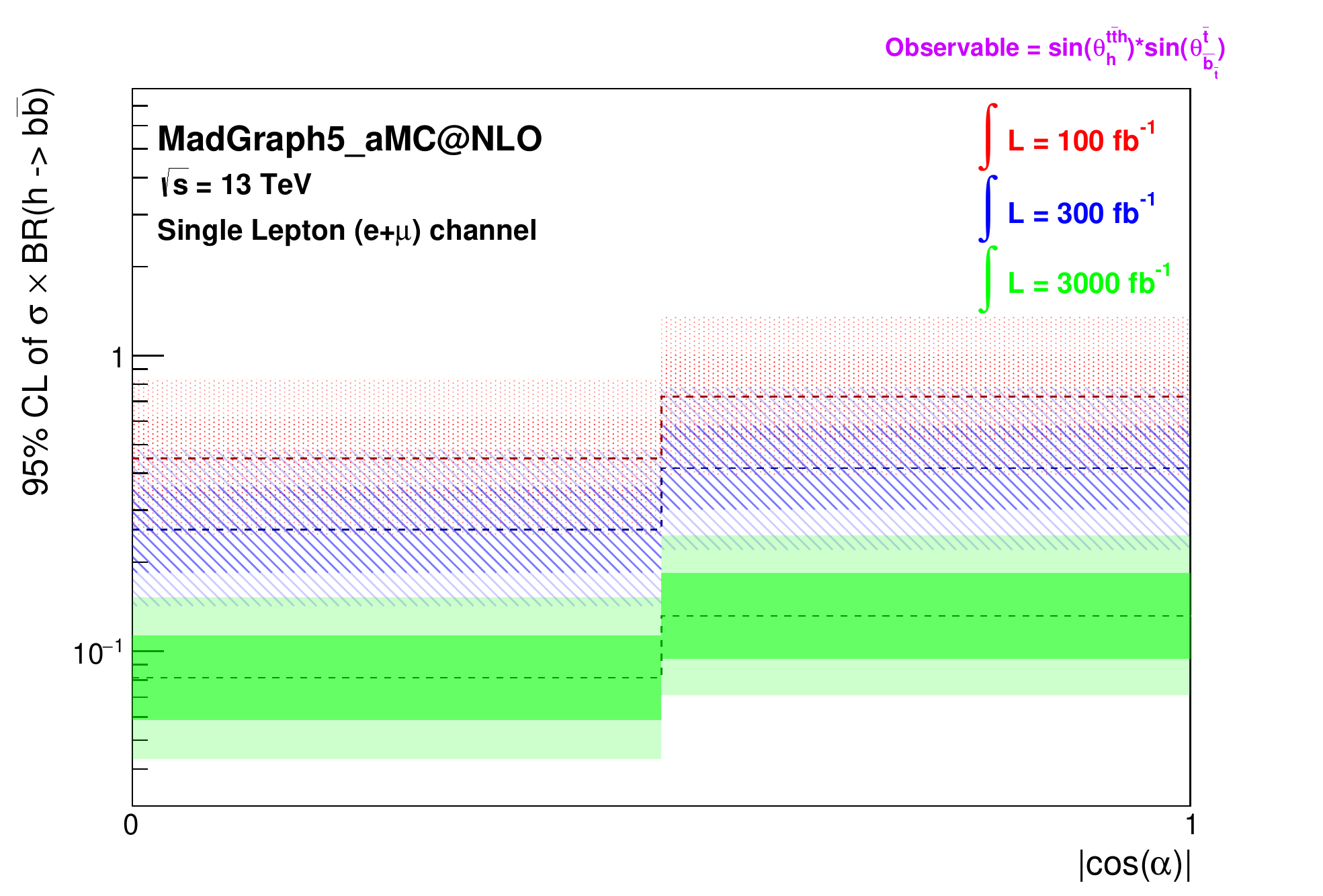,height=6.0cm,clip=} 	& \quad & \epsfig{file=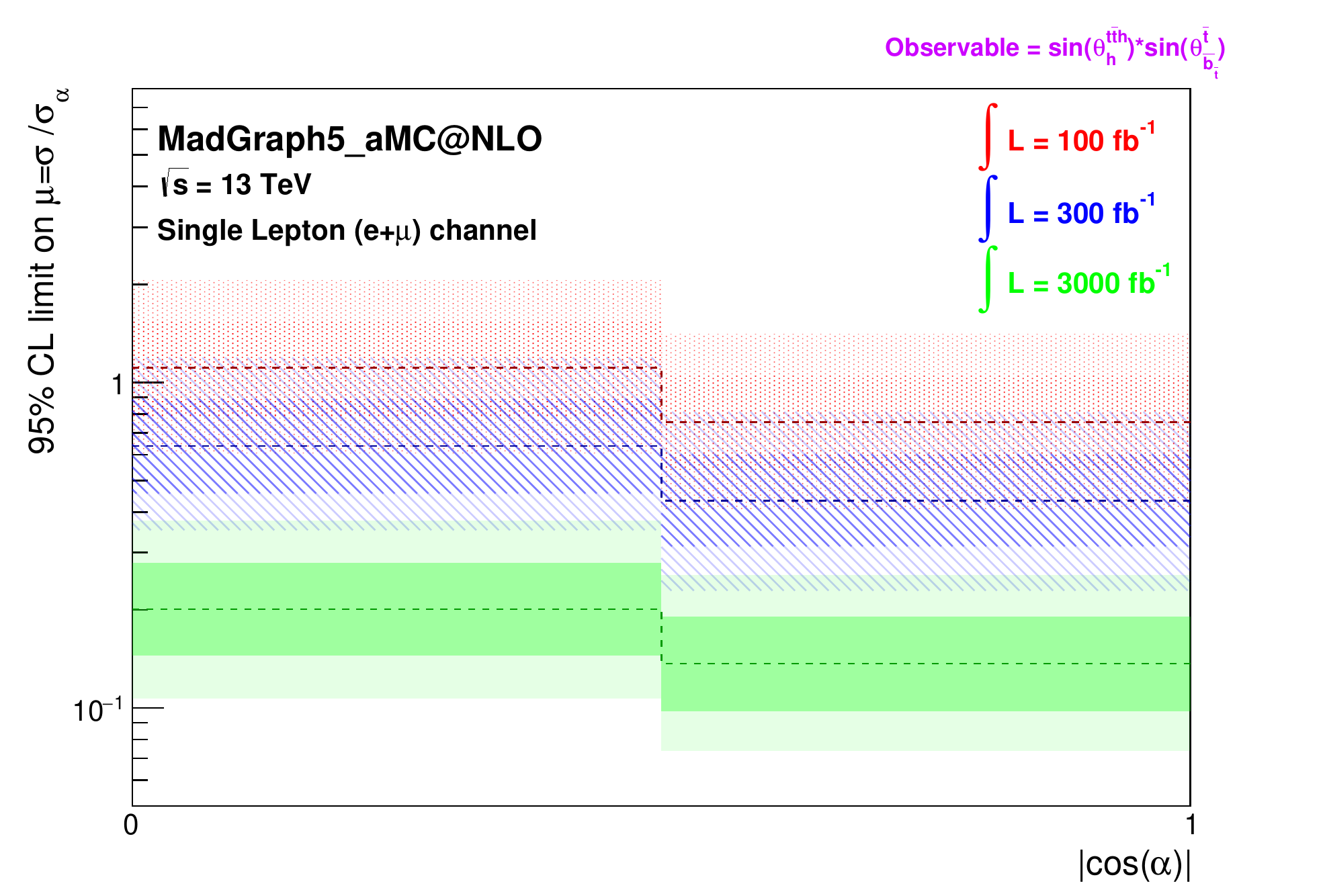,height=6.0cm,clip=}\\
\hspace*{-5mm}\epsfig{file=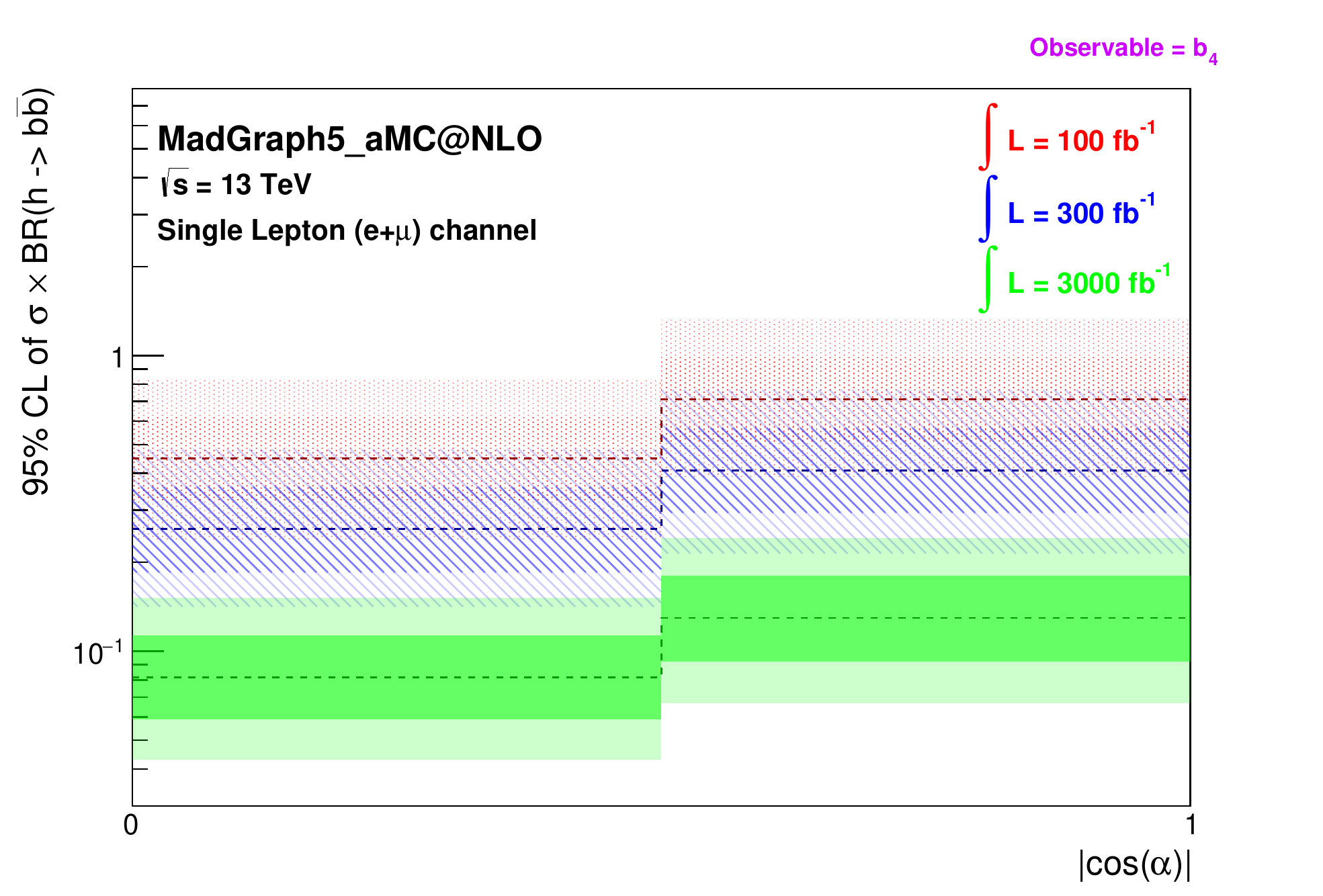,height=6.0cm,clip=} 	& \quad & \epsfig{file=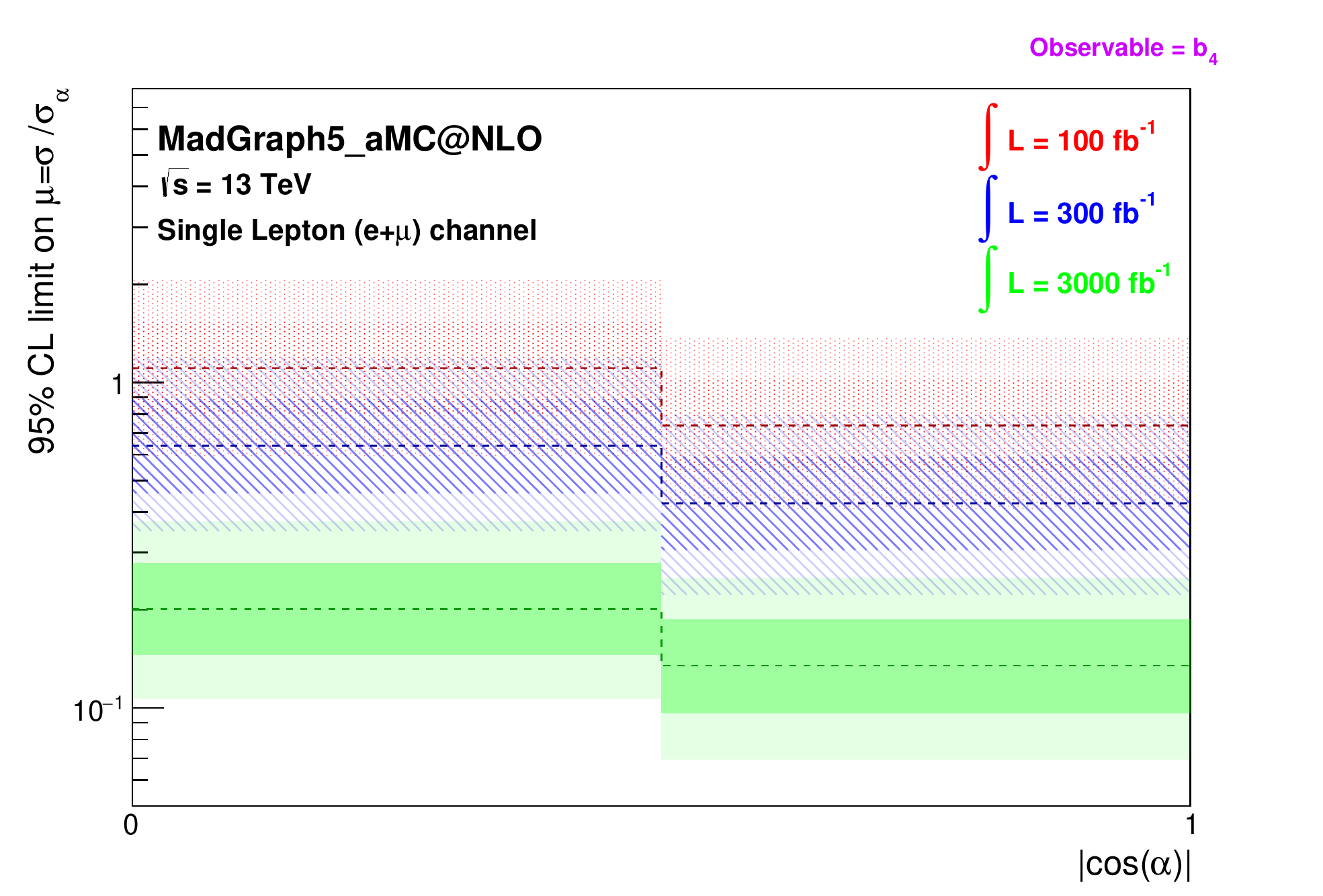,height=6.0cm,clip=}\\[-4mm]
\end{tabular}
\caption{Distributions of: (Top) 
$x_Y$=$\sin{(\theta^{t\bar{t}h}_{h})}\cos{(\theta^{\bar{t}}_{b_h})}$ (left) and 
$x_Y$=$\sin{(\theta^{t\bar{t}h}_{h})}\sin{(\theta^{\bar{t}}_{\bar{b}_{\bar{t}}})}$ (right) 
after final event selection and kinematic reconstruction at 13 TeV for 100~fb$^{-1}$ with the contributions from the full SM background and fake data; (Middle) expected limits at 95\% CL in the background-only hypothesis, for $|\cos(\alpha)|=0,1$. Limits on $\sigma\times BR(h\rightarrow b\bar{b})$ (left) and $\mu$ (right) obtained with the $\sin{(\theta^{t\bar{t}h}_{h})}\sin{(\theta^{\bar{t}}_{\bar{b}_{\bar{t}}})}$ distribution for integrated luminosities of 100, 300 and 3000~fb$^{-1}$ are shown. The lines correspond to the median, while the narrower (wider) bands correspond to the 1$\sigma$(2$\sigma$) intervals. (Bottom) The same limits as presented for the middle plots, but here for the $b_4$ distribution.}
\label{fig:NewAnal01}
\end{center}
\end{figure*}

The angular distributions can be grouped in two different categories: (i) those that exhibit similar behaviour between the scalar and pseudo-scalar signals and both different from the backgrounds and, (ii) those which are different among signals. While the first set (that may include distributions like the ones shown in Figure~\ref{fig:NewAng01} top-right or  bottom-left) is appropriate for measurements of total $t\bar t h$ production rates at the LHC which do not show strong shape dependence on the type of coupling, the second set (that may include distributions like the ones in Figure~\ref{fig:NewAng01} middle-right or bottom-right) provides sensitivity to probe the CP nature of the Higgs boson Yukawa couplings to top quarks. 
Other observables previously proposed~\citep{Artoisenet:2012st, Gunion:1996xu, tth_spin} have also been investigated. We have found that, for the semileptonic decays of $t\bar t h$ events, and after selection and kinematic reconstruction, they do not have the same sensitivity as the $b_4$ variable.
For illustration purposes, we show in Figure~\ref{fig:NewAnal01} (top) the expected number of events for the different SM backgrounds and the SM Higgs signal, after event selection and kinematic reconstruction for a luminosity of 100~fb$^{-1}$ at the LHC. Two angular distributions are shown: $x_Y$=$\sin{(\theta^{t\bar{t}h}_{h})}\cos{(\theta^{\bar{t}}_{b_h})}$ (left), and $x_Y$=$\sin{(\theta^{t\bar{t}h}_{h})}\sin{(\theta^{\bar{t}}_{\bar{b}_{\bar{t}}})}$ (right). For completeness, we also show a fake data distribution obtained by randomly sampling the expected SM signal and background distributions to mimic the intended integrated luminosity. 
The $t\bar{t}$+jets background in Figure~\ref{fig:NewAnal01} (top) includes the contributions from light and $c$-jets which, as can be seen in Table~\ref{tab:cutflow}, is a significant background after the final event selection applied in this paper. Restricting the selection to 4 $b$-tagged jets, the signal significance can increase at the expense of some statistical loss, and the background composition changes to a more $t\bar{t}b\bar{b}$ dominated sample~\cite{Aaboud:2017rss}. This is the main reason why signal angular distributions are shown against the $t\bar{t}b\bar{b}$ background.

\section{IV. Results}

Expected limits at 95\% confidence level (CL) for $\sigma\times BR(h\rightarrow b\bar{b})$ and for the signal strength, 
$\mu$\footnote{The signal strength is defined as the ratio of the measured cross section, \(\sigma \times \text{Br}\), by the SM expectation, \((\sigma \times \text{BR})_\text{SM}\), \(\mu=\frac{\sigma \times \text{Br}}{(\sigma \times \text{Br})_\text{SM}}\)} 
in the background-only hypothesis, were obtained using ROOT's TLimit~\cite{Brun:1997pa} implementation of the modified frequentist 
likelihood method (CL$_s$)~\cite{Read:2002hq,Junk:1999kv}. A test statistic was defined and computed for $10^5$ pseudo-experiments in the hypotheses of signal plus background and background only. The statistical fluctuations of the pseudo-experiments are performed with Poisson distributions. All statistical uncertainties of the expected backgrounds and signal efficiencies were taken into account in deriving the confidence level for a given signal hypothesis. The limits were calculated for the angular distribution $\sin{(\theta^{t\bar{t}h}_{h})}\sin{(\theta^{\bar{t}}_{\bar{b}_{\bar{t}}})}$ and the \(b_4\) variable. We checked that other angular distributions gave similar results. Scalar ($h=H$) and pseudo-scalar ($h=A$) signals were used, corresponding to values of the CP phase set to $|\cos(\alpha)|=\{0,1\}$ (see Equation~\ref{eq1}). Figure~\ref{fig:NewAnal01} shows the limits obtained for the angular distribution 
$\sin{(\theta^{t\bar{t}h}_{h})}\sin{(\theta^{\bar{t}}_{\bar{b}_{\bar{t}}})}$ (middle) and \(b_4\) (bottom) on the  $\sigma\times BR(h\rightarrow b\bar{b})$ (left) and signal strength $\mu$ (right). The limits were set for integrated luminosities of 100, 300 and 3000 fb$^{-1}$. 
Sensitivity to the SM $t\bar{t}H$ production with $\mu$=1 should be attained shortly after 100 fb$^{-1}$ of total integrated luminosity has been collected, using the angular distributions in this channel alone. 
The expected confidence level for the exclusion of an overall contribution to data of a pure pseudoscalar signal (A) against the SM Higgs hypothesis (H) was set at 85.5\%, 96.9\% and 100.0\% for 100, 300 and 3000~fb$^{-1}$, respectively.
The results obtained in the semileptonic channel are almost a factor 2 better than the ones presented for the dileptonic channel in~\cite{Santos:2015dja}. Combining both channels should allow to decrease the luminosity needed to probe the structure of Higgs boson couplings to the top quarks. This study, however, is outside the scope of this paper.

\section{V. Conclusions}

In this paper, we study the experimental sensitivity to the CP nature of the Higgs boson Yukawa couplings to top quarks, which can be obtained through the use of several angular observables using $t{\bar t}h$ (with $h=H,A$) events produced at the LHC. Several benchmarks for integrated luminosities were used i.e., 100, 300 and 3000~fb$^{-1}$. Semileptonic final states from $t{\bar t}h$ decays
were fully reconstructed by a kinematic fit performed with {\scshape KLFitter}. We show that, even after event selection and full kinematic reconstruction,
the shape of the new angular distributions and asymmetries is largely preserved, and can be used to discriminate between the different types of signals (scalar vs. pseudo-scalar) and the dominant irreducible SM background, $t\bar{t}b\bar{b}$. As the spin information is largely preserved, the angular distributions were used to determine expected limits at 95\% CL on $\sigma\times BR(h\rightarrow b\bar{b})$ and signal strength $\mu$. The performance obtained from the use of angular variables is compared with that of other observables commonly discussed in the literature, yielding at least the same sensitivity to the nature of the top quark Yukawa coupling, if not better. All results presented in this paper were obtained using the semileptonic final states of $t\bar{t}h$ events alone, which were found to be significantly better (around a factor 2) than the ones obtained in the dileptonic channel. Thus searches for a CP-odd component in the coupling of the Higgs boson to top quarks can be expected to improve when combining the information from both decay channels using angular observables. 

\section*{Acknowledgements}

We would like to thank S. Amor Dos Santos \textit{et al.} for providing the angular distributions from their analysis in the dileptonic channel~\cite{Santos:2015dja, AmorDosSantos:2017ayi}.

This work was partially supported by Funda\c{c}\~ao para a Ci\^encia e Tecnologia, FCT (projects CERN/FIS-NUC/0005/2015 and CERN/FP/123619/2011, grant SFRH/BPD/100379/2014 and contract IF/01589/2012/CP0180/CT0002).
Special thanks goes to our long term collaborator Filipe Veloso for the invaluable help and availability on the evaluation of the confidence limits discussed in this paper.

% ============================================================
% ============================================================
% ===========                           Bibliography                     ================= 
% ============================================================
% ============================================================

% ============================================================
% ============================================================
% ===========                               Figures                        ================= 
% ============================================================
% ============================================================
\newpage

% Figure 1

% Figure 2
\newpage

% Figure 3

% Figure 4
\newpage

% Figure 5
\newpage

\end{document}